\documentclass[trackchanges, twocolumn]{aastex701}

\shorttitle{GLASHES III}
\shortauthors{K. Morii et al.}

\usepackage{textcomp}
\usepackage{comment}
\usepackage{rotating}
\usepackage{multirow}
\usepackage{makecell}
\usepackage{subcaption}
\usepackage{amsmath}
\usepackage{enumitem}

\newcommand{\hcop}{HCO$^{+}$}
\newcommand{\hnc}{HNC}

\begin{document}

\title{Global and Local Infall in the ASHES Sample (GLASHES). III. Physical Parameters of Contracting Motion onto Dense Cores in 70 \textmu m Dark Massive Clumps} 

\author[orcid=0000-0002-6752-6061,sname=Morii, gname=Kaho]{Kaho Morii}
\altaffiliation{CfA Postdoctoral Fellow}
\affiliation{Center for Astrophysics $|$ Harvard \& Smithsonian, 60 Garden Street, Cambridge, MA 02138, USA}
\email{kaho.morii@cfa.harvard.edu}  

\author[0000-0002-7125-7685]{Patricio Sanhueza}
\affiliation{Department of Astronomy, School of Science, The University of Tokyo, 7-3-1 Hongo, Bunkyo-ku, Tokyo 113-0033, Japan}
\email{psanhueza@astron.s.u-tokyo.ac.jp} 

\author[0000-0003-2384-6589]{Qizhou Zhang}
\affiliation{Center for Astrophysics $|$ Harvard \& Smithsonian, 60 Garden Street, Cambridge, MA 02138, USA}
\email{qzhang@cfa.harvard.edu} 

\author[0000-0002-2885-1806]{Philip C. Myers}
\affiliation{Center for Astrophysics $|$ Harvard \& Smithsonian, 60 Garden Street, Cambridge, MA 02138, USA}
\email{pmyers@cfa.harvard.edu} 

\author[0000-0002-3466-6164]{James M. Jackson}
\affiliation{Green Bank Observatory, 155 Observatory Road, Green Bank, WV 24944, USA}
\email{jmjackso@nrao.edu} 

\author[0009-0007-8332-3663]{Fumiaki Yoshida}
\affiliation{Department of Astronomy, School of Science, The University of Tokyo, 7-3-1 Hongo, Bunkyo-ku, Tokyo 113-0033, Japan}
\email{fumiakiyoshida@g.ecc.u-tokyo.ac.jp} 

\begin{abstract}
Core growth is a key process in high-mass star formation. As part of the GLASHES (Global and Local Infall in the ASHES Sample) survey, we present infall velocities and mass infall rates derived from blue-asymmetric line profiles toward cores embedded in massive 70 \textmu m-dark clumps. Robust Hill5 model fits were obtained for 70 of the 135 cores exhibiting blue-asymmetric profiles, yielding infall velocities of 0.11--1.60 km s$^{-1}$ and mass infall rates ranging from $\sim$10$^{-5}$ to $5\times10^{-3}\,M_\odot\,{\rm yr}^{-1}$. Both quantities increase systematically from prestellar candidates to warm cores and outflow-associated cores.
The inferred infall velocities are close to the cores' free-fall velocities (mean $v_{\rm in}/v_{\rm ff}\sim0.75$) and $t_{\rm in}=R/v_{\rm in}$ is comparable to the free-fall time, indicating dynamical fast contraction. Hill5 infall velocity correlates moderately with the non-thermal velocity dispersion and Mach number measured from optically thin tracers, indicating that organized infall contributes to the observed line widths. The mass infall rate follows a power-law relation with core mass, $\dot{M}=1.3\times10^{-4}\,(M/M_\odot)^{0.83\pm 0.06}\,M_\odot\,{\rm yr}^{-1}$, across $0.1\,M_\odot \lesssim M \lesssim 30\,M_\odot$. The power-law index increases from $\sim$0.65 for prestellar cores to $\sim$1.10 for outflow-associated cores, likely tracing an evolution in the dominant physical process. 
The Hill5 infall velocities and mass infall rates for the GLASHES cores are larger than those of cores in nearby low-mass star forming regions but smaller than those of larger scale high-mass clumps.  
These results provide the first statistical constraints on core-scale mass accumulation during the earliest stages of high-mass star formation. 
\end{abstract}


\section{Introduction} \label{sec:intro} 
A fundamental question in star formation is how gas is accumulated from molecular-cloud scales onto individual stars. Molecular clouds exhibit a hierarchical structure consisting of clouds, clumps, filaments, and dense cores \citep[e.g.,][]{BerginTafalla07}, where dense cores are generally regarded as the immediate precursors of stars.
However, cores are not completely isolated systems, particularly in regions forming stellar clusters and high-mass stars. 
It remains actively debated whether most of the final stellar mass originates from gas initially bound within the natal core \citep[the core-fed scenario; ][]{McKeeTan03} or from material accreted from the surrounding clump on scales substantially larger than the initial core size \citep[the clump-fed scenario; ][]{Bonnell01, vazquez19, Padoan20}).

In the latter picture, an additional question concerns when the mass assembly process is most efficient. Gas accumulation may occur primarily during the prestellar core-formation phase, or it may continue throughout protostellar evolution. 
Various accretion mechanisms have been proposed, including tidal-lobe accretion, in which the gravitational potential of the parent cluster/cloud dominates the gas dynamics, and Bondi–Hoyle-like accretion, where the gravitational potential of the protostellar system itself becomes dominant \citep{BondiHoyle44,Bonnell01, Tan14}. 
Recent numerical simulations suggest that a substantial fraction of the material eventually incorporated into stars can originate well beyond the initial core boundary and may even be gravitationally unbound at early times \citep[e.g.,][]{Smith09, Padoan20, Pelkonen21, Kaalva26}.

Observationally, in high-mass star-forming regions, recent Atacama Large Millimeter/submillimeter Array (ALMA) surveys have revealed systematic increases in the average core mass, the maximum core mass, and the high-mass end of the CMF with clump evolution \citep[e.g.,][]{Pouteau22, Liu23, Louvet24, Xu24, Coletta25-almagal, Morii26, Yang26}. 
Together with the scarcity of high-mass prestellar cores \citep[$\gtrsim$30\,$M_\odot$; e.g.,][]{Zhang11, Lu15, Sanhueza17, Pillai19, Svoboda19, Sanhueza19, Morii23}, particularly in the early stages of high-mass star and cluster formation, these results support clump-fed models. However, this observational support is mostly indirect, inferring the need for core growth rather than directly detecting it.  

Direct observational searches for accretion flow or infall onto cores have mostly been conducted through velocity gradient analysis and infall line profiles. The former, velocity gradient analysis, is suitable for finding asymmetric flows along filamentary structures, which have been reported from large scales \citep[e.g.,][]{Peretto13, Imara17} down to core scales in high-mass star-forming regions \citep[e.g.,][]{Csengeri11, Henshaw14, Lu18, Chen19, Sanhueza21,Sanhueza25,Redaelli22,Zhou22, Olguin23, Olguin25, Wells24, Morii25}. 
Assuming the inclination angle of the structure, this analysis gives the accretion rate from the measured velocity gradient and the density. 

Another method uses the line profile of an optically thick line. In a centrally condensed collapsing structure, the $\tau=1$ surface of the blueshifted emission lies deeper in the object than that of the redshifted emission. As a result, the blueshifted peak traces warmer gas and becomes brighter than the redshifted peak \citep[e.g.,][]{Myers96}.
Such asymmetries have been reported over the past 30 years \citep[e.g.,][]{Mardones97, Evans99, Evans03}. In addition to many case studies in various environments, including massive protostars \citep[e.g.,][]{HoHaschick86, KetoHoHaschick88, ZhangHo97, ZhangHoOhashi98, FullerWilliamsSridharan05}, statistically significant surveys have been conducted toward low-mass cores in low-mass star-forming regions \citep[e.g.,][]{Zhou93, Tafalla98, LeeMyers11, Campbell16}, high-mass clumps \citep{Sanhueza10, Schneider10, Reiter11, Rygl13, He16, Wyrowski16, Traficante17, Jackson19, Jin-Jin21, Yang21, Xu23_hcn, Jackson26, Xie26}, and molecular clouds \citep{LeungBrown77, Myers96} using single-dish radio telescopes.
Several case studies have been carried out at core scales ($\lesssim$0.1 pc) in high-mass star-forming regions \citep[e.g.,][]{Csengeri11, Contreras18, Morii25, Gupta26}; however, the sample sizes remained too small to establish statistically robust conclusions, such as typical infall velocities, mass infall rates, or any dependence on core properties.

Building on this situation, we conducted the GLASHES (Global and Local Infall in the ASHES Sample) project using ALMA. ASHES (ALMA Survey of 70~\textmu m Dark High-mass Clumps in Early Stages) is another ALMA project aimed at revealing the initial conditions of high-mass star formation, targeting only 70~\textmu m-dark, massive, quiescent clumps with a spatial resolution of $\sim$3000--5000 au \citep{Sanhueza19, Li20, Sakai21, Morii21, Tafoya21, Sabatini22, Li22, Li23, Morii23, Izumi24, Morii24, Lin25, Morii26}.
In the GLASHES project \citep{Morii25, Morii26-glashes}, we conducted ALMA observations in Band 3 and 6 toward 31 massive clumps among the 39 ASHES targets, excluding regions hosting only low-mass cores ($<$1\,$M_\odot$), and covering $\sim$400 cores in total by single-pointing observations. We obtained \hcop\ (3--2) and \hnc\ (3--2) as optically thick tracers and N$_2$H$^+$ (1--0) as an optically thin tracer.

The main scope of this paper, following the first statistical study of these profiles \citep{Morii26-glashes}, is to derive the infall velocity and mass infall rate for the best candidates showing blue-asymmetric profiles. With the largest sample of cores embedded in high-mass clumps to date, we investigate the typical infall velocity and mass infall rate in such regions, their relation with core properties such as mass and evolutionary stage, and their differences from low-mass star-forming regions and clump-scale studies. We also investigate the underlying accretion mechanisms and the effects of infall motion on the measured line widths. 
The observational setup and the core sample are described in Sections~\ref{sec:obs} and \ref{sec:sample_select}. The analytic model used for the fitting is described in Section~\ref{sec:hill5}. Section~\ref{sec:result} presents the results of the infall-parameter measurements. We discuss their dependence on core properties, their differences from other environments, the proposed high-mass star formation scenario, and the current limitations of this study in Section~\ref{sec:discussion}. Conclusions are given in Section~\ref{sec:conclusion}.

\section{Observations and Data Reduction} \label{sec:obs} 
The observational data and reduction procedures used in this study are described in detail in \citet{Morii26-glashes}. 
Here we provide a brief summary for completeness. 
The ALMA observations were performed across the three cycles, Cycles 6 (2018.1.00299.S, PI: Y. Contreras), 10 (2023.1.01150.S, PI: K. Morii), and 11 (2024.1.01505.S, PI: K. Morii) toward 28 regions, including 4 new regions in addition to those in \citet{Morii26-glashes}. 
The data has been produced by combining the main 12-m array and the Atacama Compact Array (ACA), including both the 7-m array and the Total Power (TP) array. 
In this paper, we use \hnc\ ($J=3-2$, $\nu$=271.981111 GHz) data in Band 6. 
The detailed observational setups, such as the number of antennas, baselines, and on-source time, are summarized in Table~1 of \citet{Morii26-glashes}. 

Data reduction was carried out using CASA software package versions 6.5.4.9 and 6.6.1.17 for calibration and 6.6.3-22 for imaging \citep{CASA22}. 
After the cleaning process, we have obtained the line cubes with an angular resolution of (0.89--1.06) arcsec $\times$ (1.05--1.29) arcsec. 
The velocity resolution is 0.27 km\,s$^{-1}$. 
Average 1$\sigma$ rms noise level measured from line-free channel is $\sim$7 mJy beam$^{-1}$, corresponding to 1.0 K.  
All data shown in the paper about these three lines are ALMA 12\,m, 7\,m, and TP combined, and before the primary beam correction. 
As in \citet{Morii26-glashes}, core physical information (e.g., size, mass, density, virial parameter, and evolutionary stages) has been taken from \citet{Morii23} and \citet{Morii24} (2015.1.01539.S, 2017.1.00716.S, 2018.1.00192.S). 

\section{Sample Selection for Fitting} \label{sec:sample_select} 
\citet{Morii26-glashes} presented the complex profiles of both \hcop\ (3--2) and \hnc\ (3--2) lines and demonstrated that HNC preferentially indicates blue asymmetry profiles with less contamination by molecular outflows. 
They also classified core-averaged profiles according to their velocity difference parameter ($\delta_v$), the asymmetry parameter ($A$) and morphological features, and revealed 135 cores from 304 cores showing ``blue skew'' or ``blue double peak'' profiles in HNC with $\delta_v < 0$ and $A \gtrsim 0$ indicating blue asymmetry. 
Among them, 27 cores show multiple peaks in DCO$^+$ (3-2), N$_2$D$^+$ (3-2), DCN, (3-2) or N$_2$H$^+$ (1-0), which are used as optically thin tracers, implying the presence of multiple velocity components. 
As a result, 108 cores are selected as the best candidates for the fitting analysis, showing the blue asymmetric profiles only with a single optically-thin velocity component. 

In addition to the data used in \cite{Morii26-glashes}, follow-up observations toward another 7 regions from the ASHES project have been completed (2024.1.01505.S). 
After the same analysis as in \cite{Morii26-glashes}, 29 among 107 cores show blue asymmetry profiles, and 27 cores with a single-velocity component are added to the dataset for fitting. 

Thus, in the following analysis, we will apply the analytical model fitting to the combined 108 + 27 = 135 cores. 
\cite{Morii26-glashes} found no significant difference in core properties, such as mass, density, and evolutionary stages, between cores showing blue asymmetric profiles and those showing single-peaked or red asymmetric profiles. However, the results presented in this paper may still be biased toward more prominent or extreme cases with higher infall velocities, since our fitting is limited to cores exhibiting blue asymmetry. 

\section{Analytical Hill5 Model} \label{sec:hill5} 
The Hill5 model is an analytic radiative transfer model designed to reproduce the asymmetric, double-peaked line profiles characteristic of molecular emission from collapsing dense cores \citep[][]{De_Vries05}. 
Its name reflects its five free parameters. 
Rather than assuming discrete isothermal layers \citep[e.g.,][]{Myers96}, the model describes a core with a continuous excitation temperature gradient along the line of sight, rising from a cooler foreground value $T_0$ to a warmer peak value $T_\mathrm{peak}$ toward the core interior, superimposed with systematic infall motions.

\citet{De_Vries05} showed, by comparing several analytic radiative transfer prescriptions against a full non-LTE Monte Carlo solution, that the Hill5 model reproduces the Monte Carlo results most accurately among the analytic models tested and is the most robust against observational and instrumental uncertainties, such as limited signal-to-noise. This is mainly because the Hill5 model has an internally elevated excitation temperature, which is closer to a realistic core. This higher excitation temperature could be the result of an elevated density or an elevated kinetic temperature, or both.

At the same time, because the Hill5 model is fully analytic, the predicted line profile can be evaluated essentially instantaneously, in contrast to full non-LTE radiative transfer calculations, which require numerically solving the radiative transfer equation and are computationally far more expensive. 
This combination of improved physical realism over the two-layer model and the practicality of a closed-form solution, which makes it tractable to fit large samples of spectra with standard least-squares or Markov Chain Monte Carlo techniques, has led to the widespread adoption of the Hill5 model in infall studies across both low- and high-mass star-forming regions \citep[e.g.,][]{Campbell16, Kim21, Pillai23, Xu23_hcn, Morii25}.

The brightness temperature of the spectral line, obtained from the equation of radiative transfer, is then described as follows: 
\begin{equation}
\begin{split}
    T_B(v, \nu) &= \left[ J(T_{\rm peak}, \nu) - J(T_{\rm BG}, \nu) \right] \\
    &\times \left( \frac{1 - e^{-\tau_f(v)}}{\tau_f(v)} - e^{-\tau_f(v)} \frac{1 - e^{-\tau_r(v)}}{\tau_r(v)} \right),
\end{split}
\end{equation}
where 
\begin{align}
    \tau_f(v) = \tau \exp\left( -\frac{\{v - (v_{\rm sys, thick} + v_{\rm in})\}^2}{2\sigma^2} \right) \\
    \tau_r(v) = \tau \exp\left( -\frac{\{v -  (v_{\rm sys, thick} - v_{\rm in})\}^2}{2\sigma^2} \right) 
\end{align}
represent the front and rear sides, respectively. 
The five parameters of the model are as follows:
\begin{enumerate}[nosep]
    \item $\tau$: the central optical depth of the cloud, characterizing the overall opacity of the line. 
    \item $v_\mathrm{in}$: the infall velocity, encoding the systematic gravitational contraction.
    \item $\sigma$: the velocity dispersion, setting the width of the line profile.
    \item $v_\mathrm{sys, thick}$: the systemic velocity of the core in the local standard of rest frame.
    \item $T_\mathrm{peak}$: the peak excitation temperature reached in the interior of the core.
\end{enumerate}
We set the excitation temperature at the front of the core (the cooler foreground), $T_0$, as 2.73 K. 
Although the true value is likely higher given the warmer parent clumps in which the cores are embedded, we confirmed that $v_\mathrm{in}$, the primary focus of this paper, is not strongly affected by this choice. We therefore fixed $T_0$ at this value rather than introducing it as an additional free parameter in the fit.

A key simplifying assumption of the Hill5 model is that the infall velocity is constant and the same on the front and rear sides.  
While this assumption makes the model analytically simpler, it may not accurately reflect the true velocity structure of collapsing cores, where infall speeds generally increase with depth due to gravity. 

\section{Result} \label{sec:result} 
\subsection{Applying Hill5 Model} 
The actual fit was conducted using the affine-invariant Markov Chain Monte Carlo (MCMC) algorithm implemented in the emcee Python package \citep{Foreman-Mackey-emcee} to explore the parameter space. 
We ran 200 walkers for 10000 steps.  
For the parameter space, we used $\tau$ ranging from 0.01 to 20, $v_{\rm in}$ between 0.05 and 20 km\,s$^{-1}$, $\sigma$ between 0.05 and 2 km\,s$^{-1}$, 
and $T_{\rm peak}$ between 3 and 200 K. 

From 135 cores, fitting succeeded for 70 HNC profiles, which are averaged inside cores. 
The quality of the Hill5 fits was assessed using both the residual spectra and the reduced chi-square values. We required that the residuals, defined as the observed spectrum minus the best-fit model, showed no coherent features exceeding 3 $\sigma$ within the velocity range used for the fit, where $\sigma$ is the rms noise estimated individually for each spectrum. We also confirmed that the reduced chi-square values were of order unity, indicating that the Hill5 model provides an adequate description of the observed line profiles. 
For 14 sources with clear wings due to outflows, we applied wing-subtraction before fitting, which is explained in Appendix~\ref{sec:appendix-wing}.  
We also applied a threshold of $\tau \gtrsim$0.15, removing 5 sources for which the assumption of an optically thick regime does not hold. 
Note that failure to fit the Hill5 model does not necessarily indicate the absence of infall motions. The excluded 65 cores are statistically indistinguishable from the 70 Hill5-fitted cores in terms of mass, density, and evolutionary stage, and were excluded solely because their profiles were not well described by this simple analytic model, possibly due to more complex kinematics. 

\begin{figure*}
    \centering
    \includegraphics[width=1\linewidth]{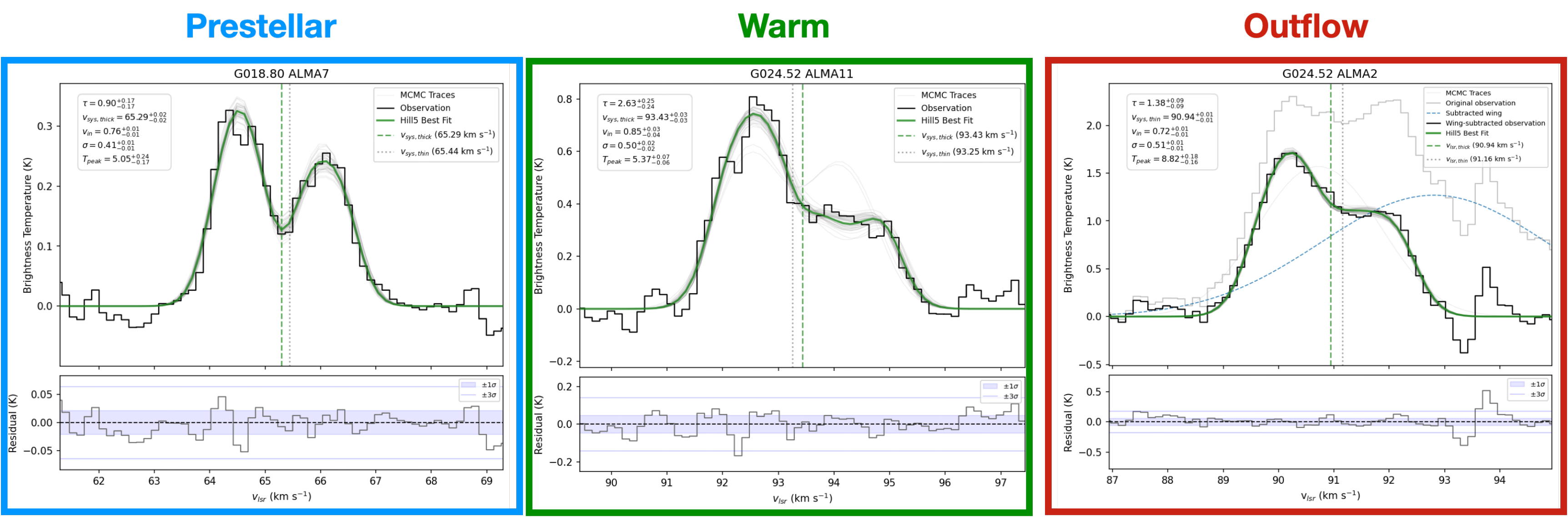}
    \caption{Examples of Hill5 model fits to the core-averaged HNC (3–2) spectra for each evolutionary category.  Black lines show the observed spectra averaged within individual dense cores, while the green curves represent the best-fit Hill5 models obtained from the MCMC analysis (the posteriors are in thin gray lines). Vertical gray dotted lines and green dashed lines indicate the systemic velocities derived from optically thin tracers ($v_{\rm sys, thin}$) and from Hill5 fitting ($v_{\rm sys, thick}$), respectively. The residuals (observed minus model) are shown in the lower panels.  The light blue dashed line in the right panel shows the subtracted broad component (see Appendix~\ref{sec:appendix-wing}).
    }
    \label{fig:ex_profiles}
\end{figure*}

Among the successfully fitted sources, 10 are associated with CO (2-1) outflows, 26 are detected with warm gas tracers warm gas tracers \citep[e.g., H$_2$CO $J_\mathrm{K_a,K_c}$=\,3$_{2,2}$--$2_{2,1}$, H$_2$CO $J_\mathrm{K_a,K_c}$=\,3$_{2,1}$--$2_{2,0}$, CH$_3$OH $J_\mathrm{K}$=\,4$_{2}$--$3_{1}$, and HC$_3$N $J$=24--23;][]{Izumi24}, and 34 are prestellar candidates without any such protostellar activities. 
They are called outflow cores, warm cores, and prestellar core candidates in this paper, and denoted as outflow, warm, and prestellar in the figures and tables, respectively. 
Furthermore, among the 70 cores analyzed, 46 (66\%) are located within filaments identified by \citet{Morii23}, while the remaining 24 (34\%) are not associated with any filament, although most of the latter are still embedded in sub-clumps \citep{Morii24}.
Figure~\ref{fig:ex_profiles} shows examples of the fitting results, and Table~\ref{tab:parameters} lists the derived parameters with their uncertainties. 

\begingroup
\renewcommand{\arraystretch}{1.15}
\begin{deluxetable*}{llccccc}
\tablecaption{Summary of Hill5 best fit parameters\tablenotemark{*} \label{tab:parameters}} 
\tablehead{
\colhead{Region} & \colhead{ID} & \colhead{$\tau$} & \colhead{$v_{\rm in}$} & \colhead{$\sigma$} & \colhead{$v_{\rm sys, thick}$} & {$T_{\rm peak}$}\\
 \colhead{} & \colhead{}  & \colhead{} & \colhead{(km\,s$^{-1}$)} & \colhead{(km\,s$^{-1}$)} & \colhead{(km\,s$^{-1}$)} & \colhead{(K)} 
}
\startdata
G016.97 &        4 & $3.38_{-0.47}^{+0.56}$ & $0.82_{-0.16}^{+0.05}$ & $0.50_{-0.04}^{+0.07}$ &  $40.99_{-0.04}^{+0.04}$ &  $4.50_{-0.07}^{+0.08}$ \\
G016.97 &       11 & $0.86_{-0.37}^{+0.40}$ & $0.54_{-0.02}^{+0.02}$ & $0.20_{-0.01}^{+0.01}$ &  $41.10_{-0.02}^{+0.02}$ &  $4.49_{-0.28}^{+0.59}$ \\
G018.80 &        7 & $0.90_{-0.17}^{+0.17}$ & $0.76_{-0.01}^{+0.01}$ & $0.41_{-0.01}^{+0.01}$ &  $65.29_{-0.02}^{+0.02}$ &  $5.05_{-0.18}^{+0.24}$ \\
G018.93 &        7 & $0.72_{-0.05}^{+0.05}$ & $0.34_{-0.00}^{+0.00}$ & $0.30_{-0.00}^{+0.00}$ &  $45.45_{-0.00}^{+0.00}$ & $10.53_{-0.28}^{+0.31}$ \\
G022.69 &        4 & $4.11_{-2.36}^{+2.45}$ & $0.20_{-0.09}^{+0.24}$ & $0.24_{-0.06}^{+0.04}$ &  $77.35_{-0.04}^{+0.04}$ &  $3.79_{-0.12}^{+0.14}$ \\
G022.69 &       10 & $1.45_{-0.58}^{+1.37}$ & $0.26_{-0.18}^{+0.02}$ & $0.20_{-0.02}^{+0.04}$ &  $76.78_{-0.02}^{+0.02}$ &  $5.02_{-0.17}^{+0.51}$ \\
G023.47 &        3 & $1.21_{-0.05}^{+0.05}$ & $1.61_{-0.00}^{+0.00}$ & $0.45_{-0.00}^{+0.00}$ &  $87.86_{-0.00}^{+0.00}$ & $10.29_{-0.14}^{+0.15}$ \\
G023.47 &       13 & $1.33_{-0.34}^{+0.37}$ & $0.46_{-0.01}^{+0.01}$ & $0.19_{-0.01}^{+0.01}$ &  $87.27_{-0.01}^{+0.01}$ &  $5.40_{-0.24}^{+0.36}$ \\
G023.47 &       17 & $2.17_{-0.15}^{+0.16}$ & $1.05_{-0.01}^{+0.01}$ & $0.41_{-0.01}^{+0.01}$ &  $88.52_{-0.01}^{+0.01}$ &  $6.42_{-0.08}^{+0.09}$ \\
G024.01 &        6 & $0.20_{-0.08}^{+0.09}$ & $0.73_{-0.00}^{+0.00}$ & $0.26_{-0.01}^{+0.01}$ &  $95.07_{-0.00}^{+0.00}$ & $21.01_{-4.30}^{+8.77}$ \\
G024.01 &        8 & $0.61_{-0.16}^{+0.16}$ & $0.86_{-0.01}^{+0.01}$ & $0.33_{-0.01}^{+0.01}$ &  $94.63_{-0.01}^{+0.01}$ &  $7.74_{-0.59}^{+0.97}$ \\
G024.52 &        2 & $1.38_{-0.09}^{+0.09}$ & $0.72_{-0.01}^{+0.01}$ & $0.51_{-0.01}^{+0.01}$ &  $90.94_{-0.01}^{+0.01}$ &  $8.82_{-0.15}^{+0.17}$ \\
G024.52 &        6 & $5.09_{-0.13}^{+0.13}$ & $0.39_{-0.01}^{+0.01}$ & $0.35_{-0.00}^{+0.00}$ &  $92.68_{-0.01}^{+0.01}$ &  $7.27_{-0.04}^{+0.04}$ \\
G024.52 &        7 & $0.91_{-0.17}^{+0.16}$ & $0.63_{-0.09}^{+0.08}$ & $0.96_{-0.04}^{+0.05}$ &  $91.92_{-0.03}^{+0.03}$ &  $5.64_{-0.14}^{+0.23}$ \\
G024.52 &        9 & $0.21_{-0.12}^{+0.20}$ & $0.10_{-0.04}^{+0.04}$ & $0.24_{-0.02}^{+0.02}$ &  $90.97_{-0.01}^{+0.01}$ & $12.66_{-3.34}^{+8.20}$ \\
G024.52 &       11 & $2.64_{-0.22}^{+0.24}$ & $0.85_{-0.04}^{+0.03}$ & $0.50_{-0.02}^{+0.02}$ &  $93.43_{-0.03}^{+0.03}$ &  $5.37_{-0.06}^{+0.07}$ \\
G024.52 &       13 & $3.07_{-0.44}^{+0.19}$ & $0.49_{-0.04}^{+0.22}$ & $0.59_{-0.07}^{+0.02}$ &  $93.13_{-0.02}^{+0.02}$ &  $6.69_{-0.09}^{+0.08}$ \\
G024.52 &       19 & $2.56_{-0.33}^{+0.47}$ & $0.73_{-0.24}^{+0.04}$ & $0.46_{-0.03}^{+0.08}$ &  $93.70_{-0.04}^{+0.04}$ &  $4.90_{-0.07}^{+0.08}$ \\
G025.16 &        8 & $2.31_{-0.10}^{+0.11}$ & $0.98_{-0.01}^{+0.01}$ & $0.49_{-0.01}^{+0.01}$ &  $64.07_{-0.01}^{+0.01}$ &  $5.48_{-0.04}^{+0.04}$ \\
G025.16 &        9 & $8.12_{-1.59}^{+2.01}$ & $0.16_{-0.04}^{+0.05}$ & $0.30_{-0.02}^{+0.02}$ &  $63.79_{-0.03}^{+0.03}$ &  $4.76_{-0.19}^{+0.17}$ \\
G025.16 &       14 & $2.87_{-0.14}^{+0.15}$ & $0.57_{-0.02}^{+0.02}$ & $0.43_{-0.01}^{+0.01}$ &  $64.11_{-0.01}^{+0.01}$ &  $6.03_{-0.04}^{+0.04}$ \\
G028.54 &        2 & $0.63_{-0.12}^{+0.12}$ & $0.65_{-0.01}^{+0.01}$ & $0.28_{-0.01}^{+0.01}$ &  $86.39_{-0.01}^{+0.01}$ &  $6.73_{-0.34}^{+0.48}$ \\
G028.54 &        4 & $0.50_{-0.09}^{+0.09}$ & $0.67_{-0.01}^{+0.01}$ & $0.37_{-0.01}^{+0.01}$ &  $86.05_{-0.01}^{+0.01}$ &  $8.33_{-0.49}^{+0.67}$ \\
G028.54 &        5 & $1.70_{-0.14}^{+0.15}$ & $0.83_{-0.01}^{+0.01}$ & $0.39_{-0.01}^{+0.01}$ &  $86.42_{-0.01}^{+0.01}$ &  $4.65_{-0.06}^{+0.06}$ \\
G028.54 &        9 & $2.17_{-0.60}^{+0.71}$ & $0.75_{-0.03}^{+0.03}$ & $0.26_{-0.02}^{+0.02}$ &  $87.29_{-0.03}^{+0.03}$ &  $4.45_{-0.15}^{+0.21}$ \\
G028.54 &       15 & $1.40_{-0.70}^{+0.40}$ & $0.36_{-0.19}^{+0.37}$ & $0.75_{-0.09}^{+0.05}$ &  $86.15_{-0.05}^{+0.05}$ &  $5.05_{-0.08}^{+0.71}$ \\
G028.56 &        2 & $3.46_{-0.15}^{+0.16}$ & $1.36_{-0.01}^{+0.01}$ & $0.32_{-0.00}^{+0.00}$ &  $88.00_{-0.01}^{+0.01}$ &  $7.06_{-0.05}^{+0.06}$ \\
G028.92 &        5 & $0.88_{-0.34}^{+0.32}$ & $0.31_{-0.09}^{+0.08}$ & $0.52_{-0.04}^{+0.06}$ &  $95.84_{-0.03}^{+0.03}$ &  $5.77_{-0.24}^{+0.63}$ \\
G030.70 &        3 & $1.09_{-0.15}^{+0.16}$ & $0.65_{-0.01}^{+0.01}$ & $0.42_{-0.01}^{+0.02}$ &  $93.40_{-0.02}^{+0.02}$ &  $5.68_{-0.16}^{+0.20}$ \\
G030.70 &        4 & $0.95_{-0.28}^{+0.26}$ & $0.48_{-0.17}^{+0.11}$ & $0.70_{-0.05}^{+0.05}$ &  $92.83_{-0.03}^{+0.03}$ &  $4.97_{-0.16}^{+0.39}$ \\
G030.70 &       13 & $1.78_{-0.39}^{+0.39}$ & $0.41_{-0.09}^{+0.12}$ & $0.53_{-0.04}^{+0.04}$ &  $93.11_{-0.04}^{+0.04}$ &  $5.47_{-0.09}^{+0.13}$ \\
G030.70 &       14 & $2.40_{-0.33}^{+0.28}$ & $0.49_{-0.07}^{+0.12}$ & $0.62_{-0.04}^{+0.03}$ &  $93.15_{-0.04}^{+0.04}$ &  $5.36_{-0.07}^{+0.07}$ \\
G030.70 &       15 & $0.61_{-0.34}^{+0.35}$ & $0.51_{-0.25}^{+0.17}$ & $0.90_{-0.09}^{+0.11}$ &  $92.92_{-0.06}^{+0.06}$ &  $5.58_{-0.44}^{+1.53}$ \\
G030.70 &       20 & $1.11_{-0.37}^{+0.34}$ & $0.35_{-0.08}^{+0.06}$ & $0.50_{-0.04}^{+0.05}$ &  $92.38_{-0.05}^{+0.05}$ &  $5.42_{-0.17}^{+0.37}$ \\
G033.33 &        8 & $0.35_{-0.11}^{+0.12}$ & $0.27_{-0.01}^{+0.01}$ & $0.24_{-0.01}^{+0.01}$ &  $93.21_{-0.01}^{+0.01}$ & $11.63_{-1.51}^{+2.74}$ \\ \\
\enddata
\tablenotetext{*}{Full version is available online as machine-readable table. }
\end{deluxetable*}
\endgroup

\begin{figure*}
    \centering
    \includegraphics[width=\linewidth]{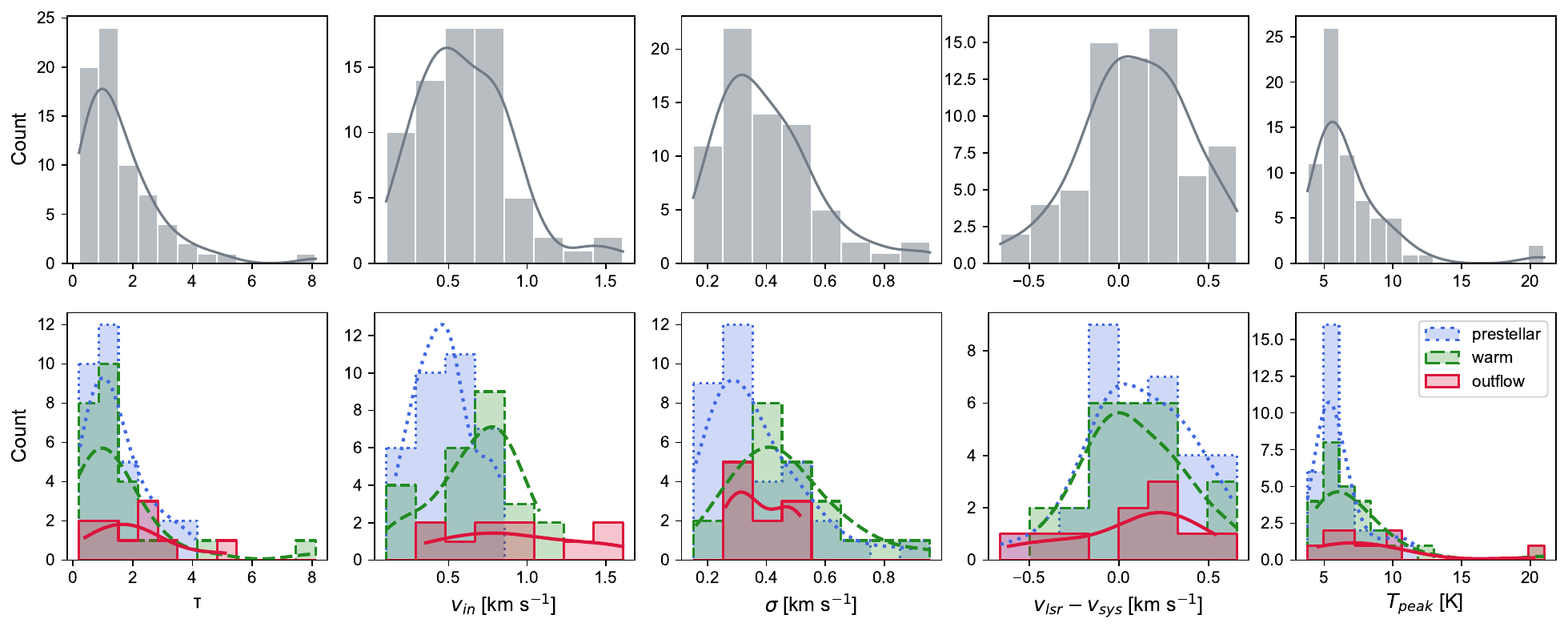}
    \caption{Distributions of the best-fit Hill5 parameters for the 70 cores with successful fits. From left to right, the panels show the optical depth ($\tau$), infall velocity ($v_{\rm in}$), velocity dispersion ($\sigma$), velocity offset from the systemic velocity ($v_{\rm sys, thick}-v_{\rm sys, thin}$), and peak excitation temperature ($T_{\rm peak}$). The upper panels show the full sample, while the lower panels separate the cores by evolutionary stage: prestellar candidates (blue, dotted), warm cores (green, dashed), and outflow-associated cores (red, solid).}
    \label{fig:hist_5para}
\end{figure*}
Figure~\ref{fig:hist_5para} shows the histograms of the estimated five parameters. The top row represents the full data set, and the bottom row shows the breakdown in terms of their evolutionary stages (prestellar candidates, warm, outflow) defined in \citet{Morii24}.  
The best-fit $\tau$ peaks around 1.5 with a tail extending to 5, and $v_{\rm in}$ varies from 0.11 to 1.60 km\,s$^{-1}$,  $\sigma$ varies from 0.15 to 0.96 km\,s$^{-1}$, and $T_{\rm peak}$ varies from 5 to 20 K. 
The fitted $v_{\rm sys, thick}$ values are consistent with the systemic velocities derived from DCO$^+$, N$_2$D$^+$, or N$_2$H$^+$ \citep{Morii26-glashes}, with offsets within $\sim$0.5 km\,s$^{-1}$. 
For some cases, best-fit $T_{\rm peak}$ is as low as 5 K, mainly due to the degeneracy with $\tau$.
However, the three velocity-related parameters ($v_{\rm in}$, $\sigma$, and $v_{\rm sys, thick}$), which are the main parameters used in this paper, are independent of this degeneracy and appear to be well constrained. 

Table~\ref{tab:parameters} lists the best-fit parameters along with their upper and lower uncertainties, which represent the $1\sigma$ (16th and 84th percentiles) confidence intervals derived from the MCMC sampling. 
The medians of each uncertainty are $d \tau=$0.18, $d v_{\rm in}=$0.014 km\,s$^{-1}$, $d \sigma=$0.013 km\,s$^{-1}$, $d T_{\rm peak}=$0.17 K, respectively, highlighting the well-constrained derived infall velocities.  

Compared to the three colors in Figure~\ref{fig:hist_5para}, cores associated with outflows (red solid lines) show somewhat different distributions from prestellar candidates (blue dotted lines) and warm cores (green dashed lines). 
Although the sample size difference and the small number of outflow cores (10) compared with prestellar (34) and warm (26) cores, they generally exhibit higher optical depths, higher $v_{\rm in}$, broader line widths, and slightly higher temperatures.  
In particular, for $v_{\rm in}$, there is a tentative trend of increase from prestellar to warm to outflow stages. 
We find no statistically significant difference in the Hill5 infall velocities between cores associated with filaments and relatively isolated cores. This may indicate that the Hill5 model is not strongly sensitive to asymmetric or non-radial flows along filaments.

\subsection{Mass Infall Rate}
\begin{figure}
    \centering
    \includegraphics[width=\linewidth]{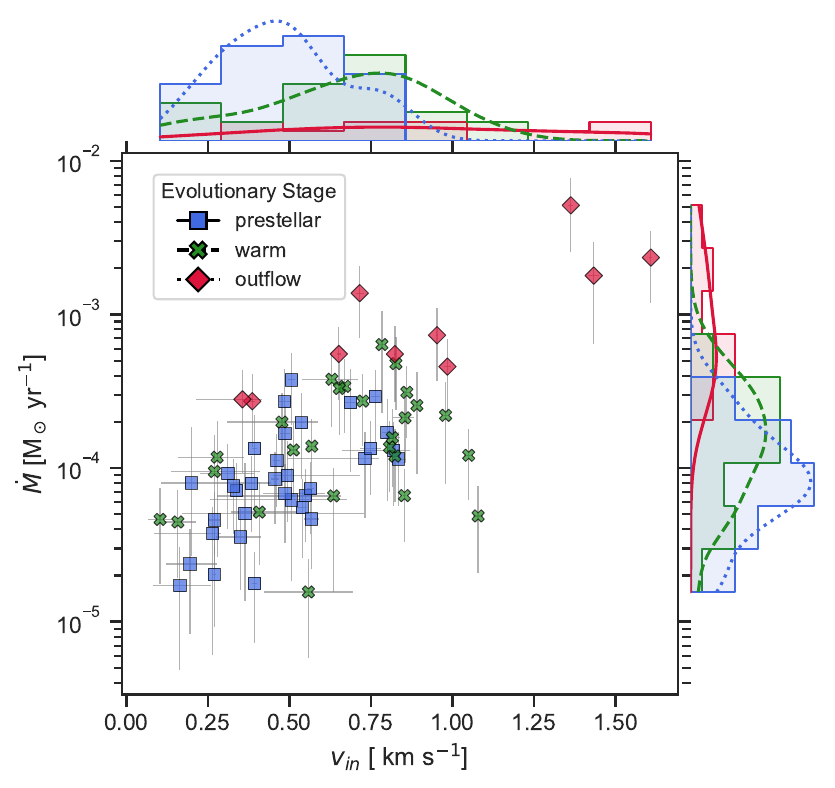}
    \caption{Mass infall rate as a function of infall velocity derived from the Hill5 model. Colors indicate the evolutionary stage of each core. 
    }
    \label{fig:Mdot_Vin_evol}
\end{figure}

The mass infall rate is then calculated from the fitted infall velocity assuming a spherically symmetric, gravity-driven infall. 
Under this assumption, the mass infall rate is given by
\begin{equation}
    \dot{M} = 4\pi R^2 \rho\, v_{\rm in},
    \label{equ:Mdot}
\end{equation} 
where $R$ is the core radius, $\rho$ is the mean gas density, and $v_{\rm in}$ is the infall velocity derived from the Hill5 model fitting. 
The core radius and mean density are taken from \citet{Morii23}, where they are estimated from 1.3 mm continuum emission.  
The derived values are in Table~\ref{tab:parameters2}. 
Figure~\ref{fig:Mdot_Vin_evol} shows the scatter plot of $v_{\rm in}$ and $\dot{M}$ color-coded by their evolutionary stages. 
$\dot{M}$ varies from 10$^{-5}$ $M_\odot$\,yr$^{-1}$ to $5\times10^{-3}$ $M_\odot$\,yr$^{-1}$ with a median of 10$^{-4}$ $M_\odot$\,yr$^{-1}$. 
Both infall velocity and mass infall rate have similar ranges as in the pilot study \citep{Morii25}. 

The uncertainty in the mass accretion rate ($\dot{M}$) was evaluated by propagating the errors of the core mass ($M$), inflow velocity ($v_{\rm in}$), and core radius ($R$). 
The uncertainty in $M$ was calculated by \citet{Morii23}, which is about 50\%. 
For the physical core radius ($R$), we adopted the relative uncertainty of distance (10\%) as the dominant source of error, since $R$ scales linearly with the distance to the region. 
Finally, the total uncertainty of $\dot{M}$ was determined via the root-sum-square of the relative uncertainties of $M$, $v_{\rm in}$, and $R$, considering the asymmetric bounds arising from the temperature and velocity errors. 

One caveat for these mass infall rate estimates is that they rely on a highly simplified assumption of spherical symmetry with a constant, characteristic speed, using the core's density and radius. 
For example, the actual infalling region may extend well beyond the core radius, where the density could be lower than the value adopted here. Perhaps more importantly, the gas distribution around these cores may not be spherically symmetric at all, given the prevalence of filamentary structures feeding into them. 

\section{Discussion} \label{sec:discussion}

\subsection{Infall Velocity and Gravitational Contraction} 
\begin{figure*}
    \centering
    \includegraphics[width=\linewidth]{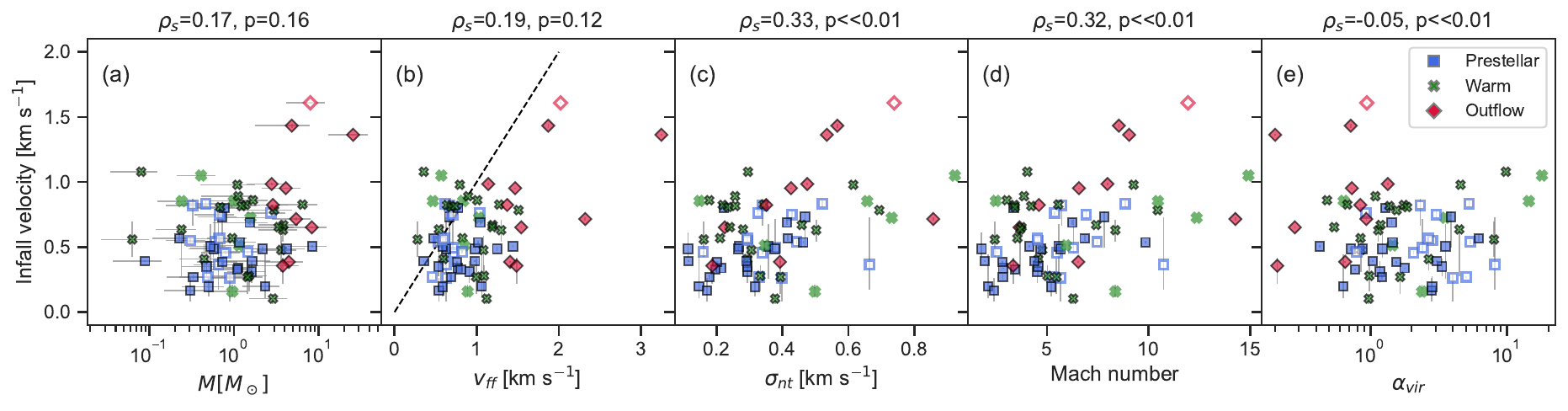}
    \caption{Hill5 infall velocity as a function of (a) core mass, (b) free-fall velocity, (c) non-thermal velocity dispersion, (d) Mach number, and (e) virial parameter. Colors indicate evolutionary stages.  The Spearman rank correlation coefficient ($\rho_s$) and p-value are indicated in each panel. Cores whose velocity dispersions are measured from N$_2$H$^+$ (1-0) are highlighted as open symbols. }
    \label{fig:Vin_phy}
\end{figure*} 
For free-fall collapse, a higher infall velocity is expected in more massive, denser objects. 
Following the case study \citep{Morii25}, we investigate whether the infall velocity has any correlation with core properties.  
Figure~\ref{fig:Vin_phy} shows the scatter plots of the Hill5 infall velocity with (a) core mass, (b) free-fall velocity of cores, (c) non-thermal velocity dispersion measured from optically thin lines, (d) Mach number ($\sigma_{nt} /c_s$), and (e) virial parameter ($\alpha_{\rm vir}$).   
All the values are in Table~\ref{tab:parameters2} in Appendix~\ref{sec:appendix-table}. 
Again, color represents its evolutionary status (prestellar, warm, and outflows are blue squares, green crosses, and red diamonds, respectively). 
The Spearman's rank correlation coefficient and p-values are denoted at the top of each panel. 

Panel (a) highlights the scatter of $v_{\rm in}$ from low-mass, prestellar plus warm cores around 0.5-1.0 km\,s$^{-1}$. 
Cores with $M\gtrsim 3 M_\odot$ or outflow-associated cores show higher $v_{\rm in}$ with an increasing trend along with core masses up to 1.5 km\,s$^{-1}$. 
Thus, the overall Spearman's correlation coefficient and its p-value rule out a simple correlation between $v_{\rm in}$ and core mass, However, there is an increase from intermediate to high-mass cores. If we estimate the Spearman's rank correlation coefficient above $3 M_\odot$, it becomes $\rho_s=$0.45. 

Panel (b) indicates the measured infall velocities are generally smaller but comparable to the cores' free-fall velocities ($v_{\rm ff} = \sqrt{2GM_{\rm core}/R_{\rm core}}$), in particular for low-mass cores. 
The measured infall velocities are generally comparable to the free-fall velocities estimated from the core masses ($v_{\rm in}/v_{\rm ff}=$0.10-3.03), with a mean ratio of 0.75, which is generally larger than cores in Perseus \citep[mean ratio of 0.44; ][]{Campbell16}. 
This suggests that the outer core envelopes are undergoing significant dynamical gravitational contraction rather than quasi-static contraction. 
Within the framework of the clump-fed scenario, such near-free-fall motions are consistent with ongoing accretion from the surrounding environment, indicating that the cores are not evolving as isolated entities but remain dynamically coupled to their parent clumps. 
The fact that a substantial fraction of the cores (25\%) exhibit $v_{\rm in}/v_{\rm ff}>1$ may reflect contributions from gravitational potentials larger than the individual core scale, although uncertainties in the simplified Hill5 model and the adopted core properties may also contribute. 

\begin{figure}
    \centering
    \includegraphics[width=0.9\linewidth]{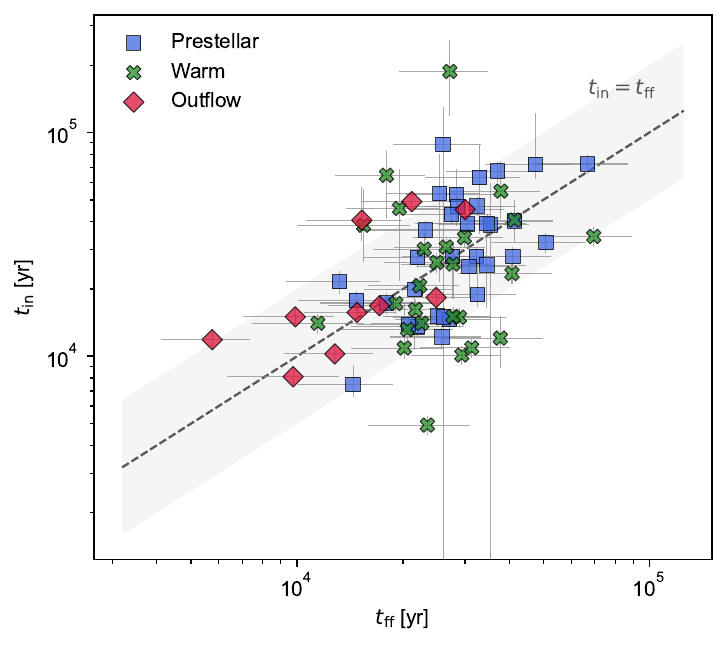}
    \caption{Scatter plot of the infall timescale, $t_{\rm in}=R/v_{\rm in}$, as a function of the free-fall timescale, $t_{\rm ff}$. Colors indicate the evolutionary stage of each core. The dashed gray line shows the one-to-one relation, $t_{\rm in}=t_{\rm ff}$, and the shaded gray region denotes agreement within a factor of 2.
    }
    \label{fig:tin_tff}
\end{figure} 
This picture is further supported by Figure~\ref{fig:tin_tff}, which compares the infall timescale, defined as $t_{\rm in}=R/v_{\rm in}$, with the free-fall timescale, $t_{\rm ff}=\sqrt{3\pi/32G\rho}$, where $G$ is the gravitational constant. 
Here, $t_{\rm in}$ represents the characteristic time for gas at radius $R$ to move inward at the measured infall velocity, $v_{\rm in}$, and $t_{\rm ff}$ represents the timescale of a global, pressure-free, spherical infall due to core self-gravity.  
Most cores lie within a factor of two of the $t_{\rm in}=t_{\rm ff}$ relation, with typical timescales of $\sim10^4$--$10^5$ yr, although a few warm cores deviate from this relation. The prestellar cores, in particular, are preferentially located close to the one-to-one relation. 

The free-fall model \citep{Hunter62} adopted here describes the gravitational contraction of a uniform sphere starting from rest (i.e., with zero initial velocity). With a nonzero initial velocity, which may be more realistic for the situation of external infall onto a core, the infall rate can become larger, as explored by \citet{Fatuzzo04}, departing from the idealized free-fall model adopted for this comparison. 
Even under such simplification, the agreement between the two timescales is consistent with the direct velocity comparison above, indicating that the observed infall motions are dynamically significant and occur on approximately the same timescale as local gravitational contraction. 
This supports a picture in which many cores, especially prestellar cores, are undergoing rapid, gravity-driven contraction, and also implies that internal pressure and magnetic field cannot be more dominant than gravity. 

\subsection{Contribution of Infall Motions to Non-thermal Linewidths} 
If the measured infall motions are associated with gravitational contraction, they may also contribute to the observed non-thermal motions. 
Panels (c) and (d) in Figure~\ref{fig:Vin_phy} show weak correlations between the Hill5 infall velocity and the non-thermal velocity dispersion and Mach number measured from DCO$^+$, N$_2$D$^+$, or N$_2$H$^+$, where the sound speed was estimated using the temperature adopted for mass estimation\footnote{Adopted temperatures are listed in Column 3 in Table~4 in \citet{Morii23}. For 19 cores, we used the rotational temperatures derived from NH$_3$ observations \citep[][]{Allingham24}. For the remaining 51 cores, we adopted the clump temperature for the core mass calculation \citep{Guzman15}. }.
The consistency with the pilot study \citep{Morii25} suggests that the observed velocity dispersion, even when derived from dense-gas tracers that are spatially compact just around continuum cores, at least partially includes local infall motions rather than purely tracing turbulent motions. 
This may lead to an overestimation of quantities that depend on the non-thermal velocity dispersion, such as the virial parameter and Mach number if the non-thermal term is assumed to be purely turbulent. 

Such a contribution of gravitationally driven motions to the observed non-thermal linewidth has long been predicted over a wide range of spatial scales, from molecular clouds and clumps down to dense cores \citep[e.g.,][]{Ballesteros-Paredes11, vazquez19, Traficante18}. In these scenarios, a significant fraction of the measured non-thermal velocity dispersion arises from global or local contraction that is observationally difficult to disentangle from turbulence. Indeed, \citet{Traficante18} argued that the observed non-thermal motions originate from both turbulence and self-gravity and that the two components cannot be observationally separated. Similarly, \citet{Ballesteros-Paredes11} suggested that the observed supersonic motions contain a dominant contracting component rather than representing purely random turbulence. 
Rotation could also contribute to the measured non-thermal velocity dispersion, although at the angular resolution of our data, we have not found clear rotation motions or velocity gradients.

We detected blue-asymmetric profiles even toward unbound cores and cores with virial parameters larger than 2 in \citet{Morii26-glashes}, and also seen in panel (e), supporting the idea that the turbulent support in virial parameters may be systematically overestimated when infall motions contribute to the measured linewidths.  
Likewise, \citet{Valeille-Manet26} reported virial parameters exceeding 2 toward protostellar cores in active high-mass star-forming regions, despite their expected gravitational instability. 
One possible explanation is that the non-thermal velocity dispersion includes a substantial contribution from gravity-driven infall, leading to an overestimate of the turbulent support. 
Another is the contribution of external pressure, which is not taken into account in our virial parameter estimation and, if included, would generally lower the derived values. 
Keeping this in mind, we find no strong correlation between Hill5 infall velocity and the virial parameter, as shown in panel (e). Regardless of the virial parameters, the range of Hill5 infall velocities of cores is similar, except for a few outflow-associated cores with smaller virial parameters. 

\subsection{Mass Infall Rate and Core Properties} \label{sec:Mdot}
The dependence of the mass infall rate on mass affects the final stellar mass. Recently, an increasing number of studies in high-mass star-forming regions have suggested the importance of external infall or core growth \citep[e.g.,][]{Nony23, Louvet24, Wells24, Xu24, Coletta25-almagal, Morii26}. 
However, the evolution of the mass infall rate and its dependence on mass, environment, and evolutionary stage have largely been unexplored due to the small sample size. 
Here, we estimate the infall velocity and mass infall rate of 70 sources embedded in similar environments, according to the clump properties extracted from a uniform data set and consistent methods, and well-characterized evolutionary stages and mass. Thus, in this section, we characterize their dependence.  

\begin{figure*}
    \centering
    \includegraphics[width=0.8\linewidth]{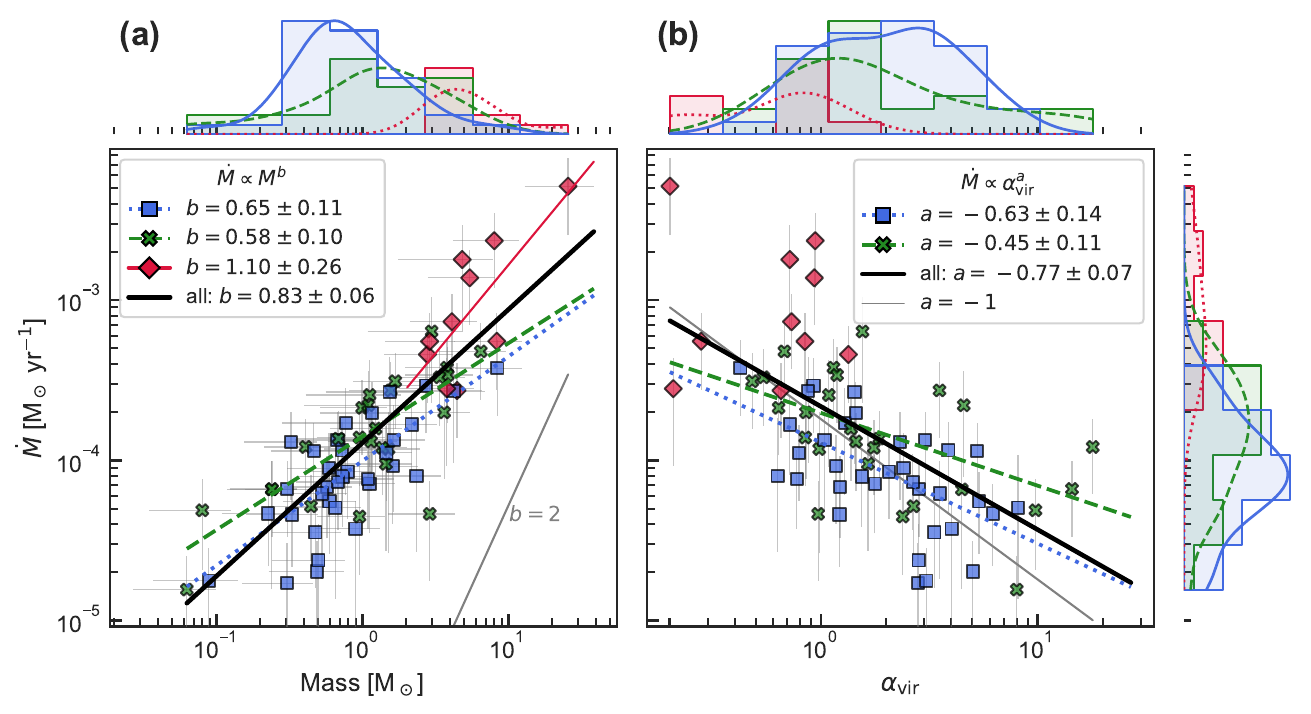}
    \caption{Mass infall rate as a function of (a) core mass and (b) virial parameters ($\alpha_{\rm vir}$). Colored lines show power-law fits of the form $\dot{M} \propto M^b$ in (a) $\dot{M} \propto \alpha_{\rm vir}^a$ in (cb, respectively. Prestellar candidates, warm cores, and outflow-associated cores are color-coded by blue, green, and red, while the black line shows the fit to the full sample. The gray line indicates the slope expected from Bondi–Hoyle accretion (b = 2) in panel (a) and a linear relation (a=-1) in panel (b). 
    }
    \label{fig:Mdot_M_avir}
\end{figure*} 
Figure~\ref{fig:Mdot_M_avir}(a) shows the $\dot{M}$--$M$ relation, together with the best-fit power-law index $b$ for $\dot{M}\propto M^b$, separately for each evolutionary stage (blue, green, and red) and for the whole sample (black). 
The best-fit indices are $0.65\pm0.11$ and $0.58\pm0.10$ for prestellar and warm cores, respectively.
These two values are similar to each other and are very close to the index expected for tidal-lobe accretion (2/3), in which the gravitational potential of the parent clump or protocluster dominates the gas dynamics; this is also consistent with another inflow study \citep{Wells24}. 
For outflow cores, in contrast, the slope steepens to $1.10\pm0.26$, intermediate between the tidal-lobe scaling and the much steeper $\dot{M}\propto M^{2}$ dependence expected when the gravitational potential of the central protostellar system itself dominates \citep[Bondi--Hoyle-like accretion;][]{Bonnell01}. 
Fitting the whole sample together ($M\sim0.1$--$30\,M_\odot$, combining both prestellar and protostellar stages) yields an intermediate index of $0.83\pm0.06$, larger than 2/3 but still much smaller than the Bondi--Hoyle value of 2. 
This evolutionary steepening, from $\sim$2/3 at the prestellar stage to $\sim$1.1 in the outflow phase, suggests a gradual transition in the dominant accretion mechanism as cores evolve: the mass dependence of the infall rate for prestellar and warm cores is consistent with accretion governed by the larger-scale gravitational potential of the clump, whereas the steeper dependence measured for outflow cores points to an increasing contribution from the protostar's own gravitational potential as it grows in mass \citep{Clark21}. 

This trend is qualitatively consistent with the picture emerging from the STARFORGE simulations \citep{Kaalva26}, in which the simple, quasi-static accretion scaling that describes lower-mass, less evolved protostars increasingly breaks down as protostars grow more massive and their accretion becomes governed by local, time-variable dynamics rather than a single global accretion law. 
We caution that the GLASHES $M$--$\dot{M}$ relation reflects a population-averaged, fixed-time comparison across cores of different mass, whereas \citet{Kaalva26} presents the time-dependent accretion history of individual protostars. The two quantities are conceptually related but not directly equivalent.

The observed scaling relation can also be compared with recent attempts to reproduce the evolution of the ASHES core mass function through core growth. 
In the CMF study within the ASHES project, \citet{Morii26} proposed a mass-growth relation of
$\dot{M}=5.33\times10^{-5}\,(M/1\,M_\odot)^{1.16}\,M_\odot\,{\rm yr}^{-1}$
to reproduce the observed transition from the prestellar CMF to the protostellar CMF. 
This relation is independent of the $\dot{M}$--$M$ relation derived from the line-profile fitting presented in this paper. 
From Figure~\ref{fig:Mdot_M_avir}(a), we obtain $\dot{M}=1.3\times10^{-4}\,(M/1\,M_\odot)^{0.83}\,M_\odot\,{\rm yr}^{-1}$ for the full sample and $\dot{M}=9.9\times10^{-5}\,(M/1\,M_\odot)^{0.65}\,M_\odot\,{\rm yr}^{-1}$ for the prestellar core candidates. 
The normalization factors agree with that of \citet{Morii26} to within a factor of a few, implying that the observed blue-asymmetric profiles support the efficient core growth rates required by clump-fed scenarios and recent CMF evolution models.

However, the power-law index, $b$, is somewhat different. 
If all cores accrete over a common timescale, a relation with $b>1$, as adopted in the CMF study, causes more massive cores to grow more efficiently and therefore tends to flatten the CMF. 
In contrast, the $b<1$ relation inferred from the infall profiles would, under the same assumption, tend to steepen the CMF. 
At first glance, this may appear inconsistent with recent observational studies reporting a flattening of the CMF from prestellar to protostellar populations and from quiescent to active star-forming environments \citep[e.g.,][]{Nony23, Louvet24, Coletta25-almagal, Morii26}. 
However, this comparison is not straightforward, because the CMF evolution model of \citet{Morii26} assumed a common accretion timescale for all cores, whereas the accretion duration may itself depend on core mass. 
Therefore, the value of $b$ alone does not uniquely determine the direction of CMF evolution. 
Furthermore, the accretion law is likely to change throughout core evolution, as shown above by the increase of the power-law index from $\sim$2/3 in the prestellar/warm phase to $\sim$1.1 in the outflow phase.
Since an index exceeding unity corresponds to a super-linear growth law under which more massive cores can pull further ahead of less massive ones, the net effect of this evolving index on the resulting CMF shape is not obvious from the present analysis alone. A larger sample spanning a broader range of evolutionary stages, combined with dedicated population-synthesis modeling, will be required to quantitatively test these possibilities.  

Figure~\ref{fig:Mdot_M_avir}(b) shows the mass infall rate as a function of the virial parameter. 
The figure shows an apparent anti-correlation, in the sense that cores with smaller virial parameters tend to have higher mass infall rates. 
However, both the mass infall rate and the virial parameter depend on $M/R$, where $M$ is the core mass and $R$ is the core radius. Therefore, such an anti-correlation is expected at least partly from their shared mathematical dependence. 
Indeed, after accounting for this co-dependence, the partial correlation coefficient does not indicate a strong independent correlation between these two quantities. 
Nevertheless, the observed trend is not simply described by a pure $\dot{M} \propto \alpha_{\rm vir}^{-1}$ relation, shown by the gray line. 
The best-fit power-law indices for the prestellar and warm cores are $-0.63$ and $-0.45$, respectively, and the index for the full sample is $-0.77$, all of which are shallower than $-1$. 
We did not fit the outflow cores separately because of their small sample size and the limited dynamic range in $\alpha_{\rm vir}$. 
Therefore, while the trend cannot be cleanly separated from the built-in parameter dependence, it is not obviously a purely mathematical artifact. 
Combining the lack of obvious correlation of Hill5 infall velocity with core mass or virial parameters (Figure~\ref{fig:Vin_phy}), the trends seen Figure~\ref{fig:Mdot_M_avir} imply that the mass infall rate is largely determined by the density or mass. As a result, more massive or denser regions with lower virial parameters are implied to be favorable sites for achieving larger infall rates and for growing more efficiently. 

\subsection{Infall Properties Compared with Other Environments} 
 \begin{figure}
    \centering
    \includegraphics[width=\linewidth]{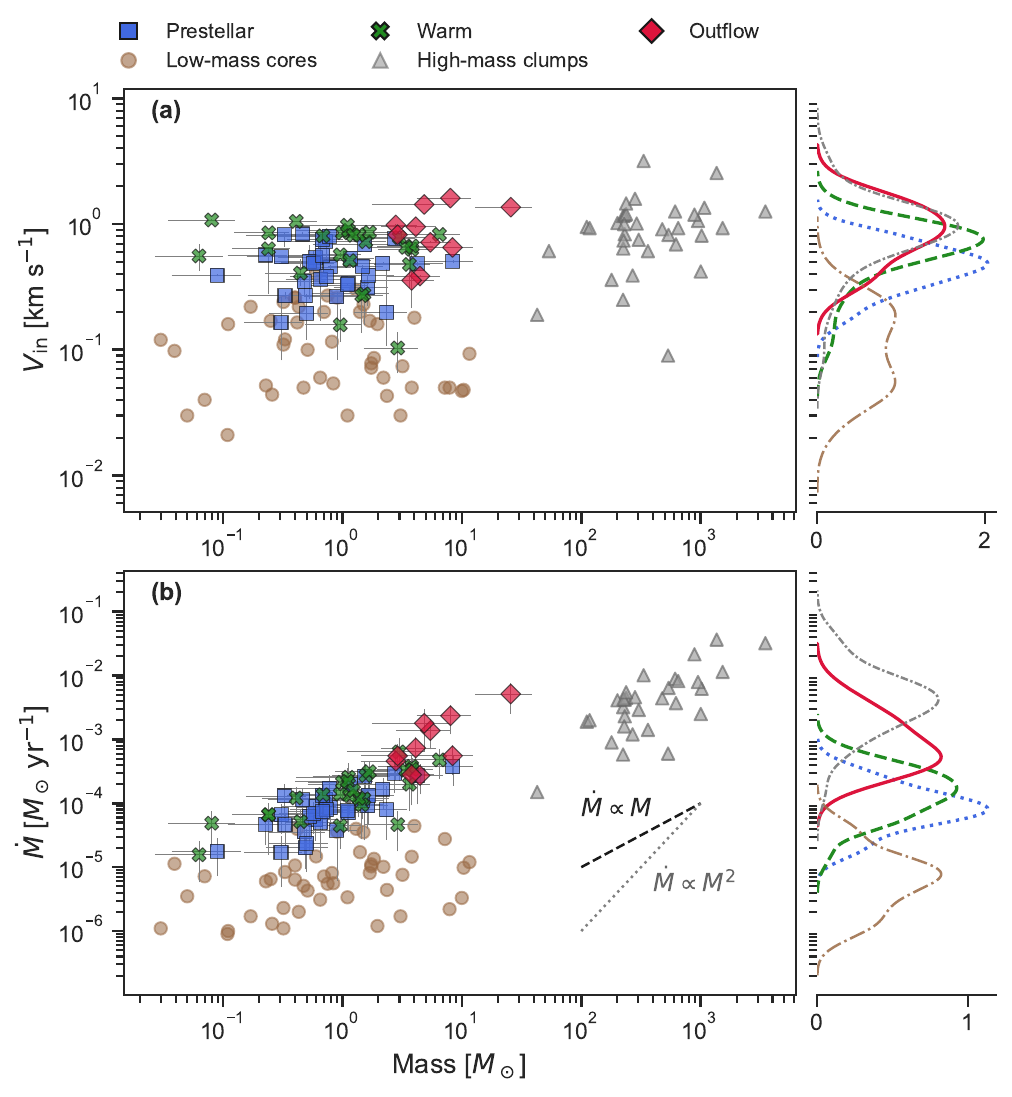}
    \caption{Comparison of the Hill5 infall velocity (panel a) and mass infall rate (panel b) as a function of core mass. Colored symbols show the GLASHES cores separated by evolutionary stage (prestellar, warm, and outflow), while brown circles and gray triangles represent measurements toward nearby low-mass dense cores and high-mass clumps reported in the literature, respectively. 
    The marginal density distributions shown on the right summarize the distributions of infall properties for each population. The dashed and dotted lines in panel (b) indicate $\dot{M}\propto M$ and $\dot{M}\propto M^2$, respectively. 
    GLASHES cores occupy an intermediate regime between nearby low-mass cores and high-mass clumps, bridging core-scale and clump-scale contraction in high-mass star-forming regions.
    }
    \label{fig:Comp_ref}
\end{figure} 
Figure~\ref{fig:Comp_ref} places the GLASHES cores in the broader context of infall studies spanning low-mass dense cores in nearby low-mass star forming regions \citep{Lee01, Campbell16, Kim21} and high-mass clumps \citep{Fuller05, Klaassen12, Rygl13, Pillai23, Xu23_hcn}. 
Most studies of high-mass clumps employed two-layer models for HCO$^+$ (1--0), except for \citet[][HCN (4--3)]{Xu23_hcn} and \citet[][CO (4--3)]{Yue21}. 
In nearby low-mass star-forming regions, infall velocities were estimated using a two-layer model for CS (2--1) \citep{Lee01} or Hill5 models for HCN (1--0) \citep{Kim21} and HCO$^+$ (3--2) \citep{Campbell16}. 
All of these previous observations were conducted with single-dish telescopes.

In the $M$--$V_{\rm in}$ plane, the GLASHES cores occupy a region intermediate between nearby low-mass dense cores and high-mass clumps. 
At a given core mass, the measured infall velocities are systematically larger than those typically observed in nearby low-mass cores, while remaining slightly below the values commonly reported for high-mass clumps. 
A clearer difference is found in the $M$--$\dot{M}$ plot.  
The GLASHES cores exhibit higher mass infall rates than nearby low-mass cores measured on comparable spatial scales, yet remain below the clump-scale accretion rates observed in massive star-forming regions. 

Quantitatively, the infall velocities reported toward low-mass cores are typically subsonic to transonic, i.e., of order the sound speed $c_s$, consistent with the rate predicted by the self-similar, inside-out collapse solution of \citet{Shu87}, $\dot{M} \sim c_s^3/G \sim 10^{-6}\,M_\odot\,{\rm yr^{-1}}$ for a typical core temperature of $\sim$10~K \citep[e.g.,][]{Mardones97, Tafalla98, Evans99, LeeMyers11, Campbell16}.   
The infall velocities and mass infall rates measured toward the GLASHES cores, in contrast, are systematically larger by two to three orders of magnitude than these low-mass values. 
The higher infall velocities and mass infall rates relative to nearby low-mass cores imply that core-scale infall motion in IRDCs proceeds within environments characterized by larger gas reservoirs, higher densities, and deeper gravitational potentials. 

The high-mass clump measurements, typically associated with clump masses of $\sim10^3\,M_\odot$, appear broadly consistent with an extrapolation of the observed $\dot{M}$--$M$ trend (see the dashed and dotted lines in the plot). 
Although the measurements probe different spatial scales and employ different molecular tracers, the continuity across nearly three orders of magnitude in mass suggests a connection between clump-scale infall and core growth. 
This behavior is qualitatively consistent with clump-fed models, in which dense cores are continuously replenished by material supplied from their parental clumps. 

A more direct test of this picture would be a comparison between the core-scale and clump-scale infall rates within the same sources. Several of the 31 parent clumps in our sample also exhibit blue-asymmetric line profiles \citep[Section 4.3 of][]{Morii26-glashes}, and we attempted Hill5 fits to them as well. However, the available 3--2 transitions toward the clumps are close to single-peaked with only slight asymmetry, rather than exhibiting the clearer double-peaked structure required for a reliable Hill5 fit, and we were therefore unable to obtain robust clump-scale infall velocities or mass infall rates from the current data. Observations of lower-$J$ transitions, better suited to tracing the more extended and lower-density clump-scale gas, would be needed to directly test whether the clump-scale infall rate correlates with the accretion rate onto the embedded cores. 

\subsection{Proposed Scenario for High-mass Star Formation in Cluster-forming Clumps}
The mass infall properties revealed in this work, combined with the core and clump properties characterized throughout the ASHES project, offer new constraints on how high-mass stars form, in particular, which environments are capable of producing high-mass stars and what kinds of structures are likely their progenitors.

The ASHES survey has revealed numerous dense fragments already present in 70\,$\mu$m-dark massive clumps, candidates for the earliest stages of high-mass star formation \citep{Sanhueza19, Morii23}, with protostellar activity already underway as traced by molecular outflows \citep{Li20, Tafoya21,Morii21, Sakai21,Izumi24, Lin25}. 
Among the 39 clumps in the survey, 10 show no outflow signatures, and three of these also lack warm-gas detections. The maximum core masses in these least evolved systems fall mostly within a factor of two of the thermal Jeans mass of their host clumps ($\sim$2--5\,$M_\odot$; \citealt{Morii24}). 
Yet as shown in panel~(a) of Figure~\ref{fig:Mdot_M_avir}, even these low-mass, prestellar-stage cores are inferred to be experiencing mass infall rates of $\gtrsim$10$^{-4}$\,$M_\odot$\,yr$^{-1}$. 
Moreover, as seen in Figures~\ref{fig:Vin_phy} and \ref{fig:tin_tff}, Hill5 infall velocities around prestellar cores are already close to their free-fall velocities. 
Sustained over the typical outflow dynamical timescale of $\sim$10$^4$\,yr measured in the ASHES cores \citep{Li20}, such infall rates are sufficient for a core to accumulate $\gtrsim$10\,$M_\odot$, consistent with the typical masses of the outflow-associated cores in our sample \citep{Morii26}. 
Once a core reaches $\gtrsim$a few $\times$10\,$M_\odot$, the mass infall rate is expected to rise further to $\gtrsim$10$^{-3}$\,$M_\odot$\,yr$^{-1}$, as implied by the $\dot{M}$--$M$ scaling relation derived in Section~\ref{sec:Mdot}, likely following the tidal-lobe accretion. 
Both the infall timescale and the free-fall timescale are typically $\sim10^4$--$10^5$\,yr. Thus, even though the observed infall proceeds at a speed close to free-fall, it can persist long enough for substantial core growth. In fact, if the mass infall rate increases with time, the masses grow even faster. 
Thus, if the inferred mass infall rates are maintained over these timescales, initially low-mass prestellar cores can accrete enough material to become high-mass cores. 

A key physical point emerging from Figure~\ref{fig:Vin_phy} is that the Hill5 infall velocity itself shows no statistically significant dependence on core mass or virial parameter (Figure~\ref{fig:Vin_phy}a and e), whereas the mass infall rate shows a clear increase (Figure~\ref{fig:Mdot_M_avir}). 
Since $\dot{M} = 4\pi R^2 \rho\, v_{\rm in}$, this implies that the higher mass infall rates seen around more massive and more evolved cores are driven primarily by their larger density or mass rather than by a systematic increase in the infall velocity itself.

\begin{figure}
    \centering
    \includegraphics[width=0.89\linewidth]{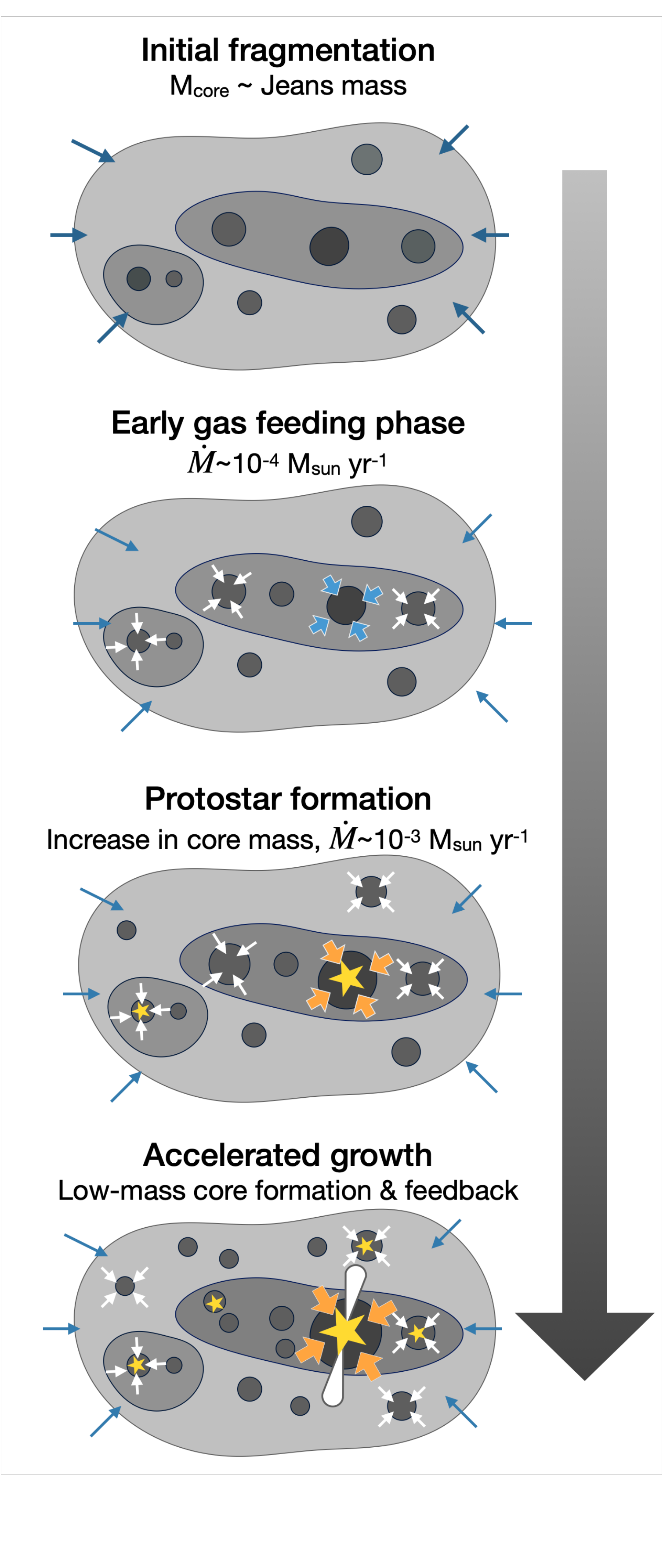}
     \caption{Schematic illustration of the proposed formation scenario. The size of each circle represents the relative core mass, and the shade of gray represents the relative core density. The blue arrows at the outer edge of the clump indicate large-scale contraction of the clump, while the white arrows around individual cores indicate gas feeding onto each core. The thick light-blue and orange arrows indicate enhanced gas feeding onto the core that eventually forms a high-mass star, and the star symbols indicate the formation of a protostar within that core. The white lobe in the bottom panel represents a molecular outflow launched from the high-mass protostellar core.}
    \label{fig:scenario}
\end{figure}
Based on these observational results, we propose the following scenario for high-mass star formation in 70\,$\mu$m-dark massive clumps (see Figure~\ref{fig:scenario}). 
In the earliest stage, the clump fragments into multiple density enhancements with masses of the order of the local thermal Jeans mass. Among them, the denser and more massive seeds, embedded in higher density gas, are the most likely seeds of future high-mass stars. 
These seed cores are not yet high-mass cores, but they can begin to grow immediately through external infall from the surrounding clump with a velocity comparable to the free-fall velocity and at rates of $\gtrsim 10^{-4}\,M_\odot\,{\rm yr}^{-1}$. As the core gains mass, the infall rate is expected to increase, producing accelerated growth. A protostar can therefore form while the core is still of low or intermediate mass, and the core subsequently becomes a genuinely high-mass protostellar core through continued mass feeding.  

In this picture, within a single cluster-forming clump, cores that eventually form high-mass stars are not distinguished from those forming low-mass stars by being born as high-mass prestellar cores. Instead, they represent the subset of early, relatively dense and massive seed cores that subsequently experience both higher infall rates and longer-lasting external accretion. The longer available growth time is partly a consequence of their early formation, while the higher infall rate likely reflects their larger core mass, higher density, and more substantial surrounding gas reservoir.  
Considering the destructive effects of turbulence and feedback, which may disperse cores \citep[e.g.,][]{Kaalva26}, continued growth may additionally require either sufficiently rapid growth before feedback becomes effective, or a protected location shielded from such disruption.

The present ASHES results show no clear preferred geometric location for these seeds on the sub-parsec scales resolved by ALMA. The most massive cores are not preferentially located at the small-scale hub-like structures identified within the clumps, nor exactly at the clump peak or cluster center in the 70 $\mu$m-dark stage \citep{Morii23}. The revealed initial core distribution is sub-clustered rather than centrally concentrated \citep{Morii23}. 
However, this does not mean that location is irrelevant on larger scales. They still reside within the inner $\sim$1 pc of dense massive clumps, where the column density is already enhanced. Therefore, although the current observations do not identify a unique geometric location, such as a small-scale hub-like structure or the exact center of the clump, as a necessary condition for enhanced infall, such locations may still be favorable when they coincide with locally dense gas, a large surrounding reservoir, or a deep gravitational potential, suggesting that the relevant factor is the local physical environment rather than the geometric label alone. 

In the cluster-forming clump environment, fragmentation and accretion proceed simultaneously. As the clump continues to fragment, new low-mass density enhancements are also produced and may grow modestly while forming the numerous low-mass protostars. At the same time, the earliest and relatively densest seed cores have a head start in mass growth and can accrete at higher rates, so that they progressively increase the maximum core mass in the region. 
Thus, as the cluster-forming clump evolves, both the number of low-mass cores and the upper end of the core-mass distribution increase, while low-mass cores remain the dominant population. 
In terms of the CMF, this corresponds to a distribution whose low-mass end continues to be replenished by newly formed fragments, while its high-mass end is built up by the continued growth of the earliest, densest seeds. 
This is broadly consistent with the observed absence of high-mass prestellar cores in IRDCs, because high-mass stars in this picture have no need to originate from an initially high-mass prestellar core, but can instead form from low- or intermediate-mass seeds that grow in situ. 
This picture is also consistent with recent ALMA surveys that study the evolution of core properties across cluster-forming regions, and report increases in the maximum core mass and the number of fragments with cluster evolution \citep[e.g.,][]{Pouteau22, Traficante23, Xu24-assemble, Coletta25-almagal}. 

Finally, the absence or rarity of high-mass star formation in nearby, lower-mass star-forming regions can be understood naturally within this scenario: lower clump masses imply smaller velocity dispersions, shallower gravitational potentials, and smaller gas reservoirs, all of which result in substantially lower mass infall rates that are insufficient to grow cores to high masses within a core free-fall time (10$^4$--10$^5$\,yr). 

\subsection{Current Limitations} 
This study presents the largest sample of core-scale infall velocities derived from Hill5 model fitting in massive clumps to date. 
The measurements were obtained using a uniform analysis procedure and a homogeneous observational setup, enabling a statistical comparison across a large sample of dense cores. 
Nevertheless, several limitations should be considered when interpreting the results and extrapolating them to the broader population of high-mass star-forming regions.

First, our sample is subject to an evolutionary bias. 
All targets are located within 70 \textmu m-dark clumps, which are thought to represent the earliest phases of high-mass star formation \citep{Rathborne06,Chambers09,Sanhueza12} and are therefore well suited for probing the initial conditions of collapse \citep{Sanhueza13,tan13}. 
One advantage of this selection is that the line profiles are less affected by strong stellar feedback, which can easily obscure infall signatures and even suppress star formation, even though 70 \textmu m-dark clumps are not completely quiescent and often already contain multiple fragments and molecular outflows \citep[e.g.,][]{Svoboda19, Sanhueza19, Li20, Morii21, Traficante23, Morii24}. 
Because our sample excludes more evolved high-mass star-forming regions, the infall properties may differ substantially from those derived here at later evolutionary stages, as clumps contract and embedded cores grow. 

Second, the inferred infall velocities depend on the adopted model. 
The Hill5 model assumes a simple velocity structure with a constant infall velocity and a monotonic increase in excitation temperature toward the core center. 
The Hill5 infall velocity is a characteristic speed needed to match line profiles of infall asymmetry, and not based on a gravitational model. 
In this work, we adopted this model to obtain first-order estimates and to facilitate comparison with previous studies. 
However, more realistic velocity structures, such as radial velocity gradients or multiple kinematic components, may lead to systematically different infall velocity estimates. 
Such detailed modeling may provide a better-constrained estimate of the infall timescale or core-growth duration time. 
Constraining such models would require higher-quality data and/or multiple molecular tracers. In addition, our current angular resolution is insufficient to directly determine the internal density profiles of individual cores. 
Therefore, quantities such as the mass infall rate remain subject to uncertainties associated with the assumed density structure.

Third, the sample of fitted sources is affected by a detection bias. 
Although this survey contains a large number of dense cores, reliable Hill5 fits were obtained for only 70 objects. 
Complex large-scale motions, line blending, projection effects, and contamination from outflows can all hinder the identification of clear blue-asymmetric profiles. As a consequence, the cores with measurable infall velocities may not fully represent the underlying core population. 
Some cores associated with outflows were fitted after subtracting wing/tail emission, which introduces additional uncertainty into the derived infall properties. 
Furthermore, the most massive core with a successful infall measurement has a mass of  $\sim$30 $M_\odot$, while the majority of fitted cores have masses below 10 $M_\odot$. 
Our conclusions are therefore primarily constrained by the low- and intermediate-mass core population embedded within massive clumps at the earliest stages of evolution. 

Finally, uncertainties associated with geometry and projection effects remain difficult to quantify. Numerical simulations suggest that infall signatures can be significantly affected by complex gas motions and viewing-angle effects \citep{Smith12, Smith13}. 
Furthermore, the true geometry of the accretion flow remains uncertain. 
While some cores may accrete material through approximately spherical contraction, others may be fed preferentially along filamentary structures with a smaller solid angle. 
Therefore, the inferred infall velocities and mass infall rates should be regarded as effective quantities derived under a common set of simplifying assumptions. 
Despite these uncertainties, such estimates provide a useful and internally consistent framework for characterizing and comparing infall properties across a large sample of cores.

Overall, these limitations should be kept in mind when interpreting the quantitative values presented in this work. Future observations with higher angular resolution, improved sensitivity, and multiple molecular tracers will be essential for establishing a more complete picture of core-scale accretion in massive star-forming regions.

\section{Conclusion} \label{sec:conclusion}
We have conducted a statistical analysis of infall motions toward dense cores embedded in 70\,\textmu m-dark massive clumps from the ASHES project, using the asymmetric, double-peaked HNC (3--2) line profiles as signatures of contracting motions. Our main results are summarized as follows:

\begin{enumerate}
\item From the full survey, we identify 135 cores exhibiting blue-skew or blue-double-peaked HNC profiles indicative of infall. Fitting these profiles with the analytic Hill5 model via MCMC, we obtain robust fits for 70 cores spanning prestellar, warm, and outflow-associated evolutionary stages.

\item The fitted infall velocities range from 0.11 to 1.60~km~s$^{-1}$, and the resulting mass infall rates range from $10^{-5}$ to $5\times10^{-3} \,M_\odot\,{\rm yr}^{-1}$ (median $\sim$10$^{-4}\,M_\odot\,{\rm yr}^{-1}$). Both quantities tend to increase from prestellar candidates through warm cores to outflow-associated cores.

\item The measured infall velocities are comparable to the cores' free-fall velocities, with a mean ratio of $v_{\rm in}/v_{\rm ff}\approx0.75$, and $t_{\rm in}=R/v_{\rm in}$ is comparable to free-fall time, indicating that the outer envelopes of these cores are undergoing dynamically significant gravitational contraction. This is consistent with ongoing accretion from the surrounding clump material, rather than the evolution of cores as isolated, gravitationally detached entities.

\item Hill5 infall velocity correlates moderately with the non-thermal velocity dispersion and Mach number derived from optically thin dense-gas tracers. 
This supports the idea, consistent with theoretical expectations and a pilot case study, that the observed non-thermal linewidths in dense cores include a non-negligible contribution from organized infall motions rather than turbulence alone.

\item The mass infall rate scales with core mass as $\dot{M}\propto M^{0.83\pm0.06}$ across the full sample ($0.1$--$30\,M_\odot$). The power-law index is shallower for prestellar candidates ($0.65\pm0.11$), consistent with tidal-lobe accretion, and steeper for outflow-associated cores ($1.10\pm0.26$), suggestive of a transition toward a different dominant accretion mode,  potentially Bondi-Hoyle-like accretion,  as cores evolve and increasingly dominate their local gravitational potential.

\item Placed in the context of previous infall studies, the GLASHES cores show systematically higher Hill5 infall velocities and mass infall rates than nearby low-mass dense cores, yet remain below the clump-scale inflow rates measured in more evolved high-mass star-forming regions. This intermediate position supports a hierarchical, clump-fed picture in which core-scale accretion is continuously replenished by larger-scale clump infall. 
\end{enumerate}

Together, these results demonstrate that 70\,\textmu m-dark clumps, representing the earliest, least feedback-affected stages of high-mass star formation, already exhibit organized infall motions whose strength scales systematically with core mass and evolutionary stage. This sample, the largest uniform set of core-scale infall measurements obtained to date in such clumps, provides new statistical constraints on how mass accretion proceeds in the earliest phases of high-mass star and cluster formation, and offers a benchmark against which future, higher-resolution observations and more detailed infall models can be compared.

\begin{acknowledgments}
We thank the anonymous referee for their careful reading and constructive comments, which improved the manuscript. 
PS was partially supported by a Grant-in-Aid for Scientific Research (KAKENHI Number JP26H02066 and JP23H01221) of JSPS. 
This paper makes use of the following ALMA data: ADS/JAO.ALMA\#2015.1.01539.S (PI: P. Sanhueza), ADS/JAO.ALMA\#2017.1.00716.S (PI: P. Sanhueza), ADS/JAO.ALMA\#2018.1.00192.S (PI: P. Sanhueza), ADS/JAO.ALMA\#2018.1.00299.S (PI: Y. Contreras), ADS/JAO.ALMA\#2023.1.01150.S (PI: K. Morii) and ADS/JAO.ALMA\#2024.1.01505.S (PI: K. Morii). 
ALMA is a partnership of ESO (representing its member states), NSF (USA) and NINS (Japan), together with NRC (Canada), NSTC and ASIAA (Taiwan), and KASI (Republic of Korea), in cooperation with the Republic of Chile. The Joint ALMA Observatory is operated by ESO, AUI/NRAO and NAOJ. 
The National Radio Astronomy Observatory is a facility of the National Science Foundation operated under cooperative agreement by Associated Universities, Inc.
\end{acknowledgments}

\facility {ALMA} 
\software{Astropy \citep{Astropy22}, CASA \citep{CASA22},  Matplotlib \citep{Hunter_matplotlib}, Numpy \citep{Harris20_numpy}, seaborn \citep{Waskom2021_seaborn}, emcee \citep{Foreman-Mackey-emcee}. } 

\appendix
\section{Outflow-component Subtraction} \label{sec:appendix-wing}
\begin{figure}
    \centering
    \includegraphics[width=0.8\linewidth]{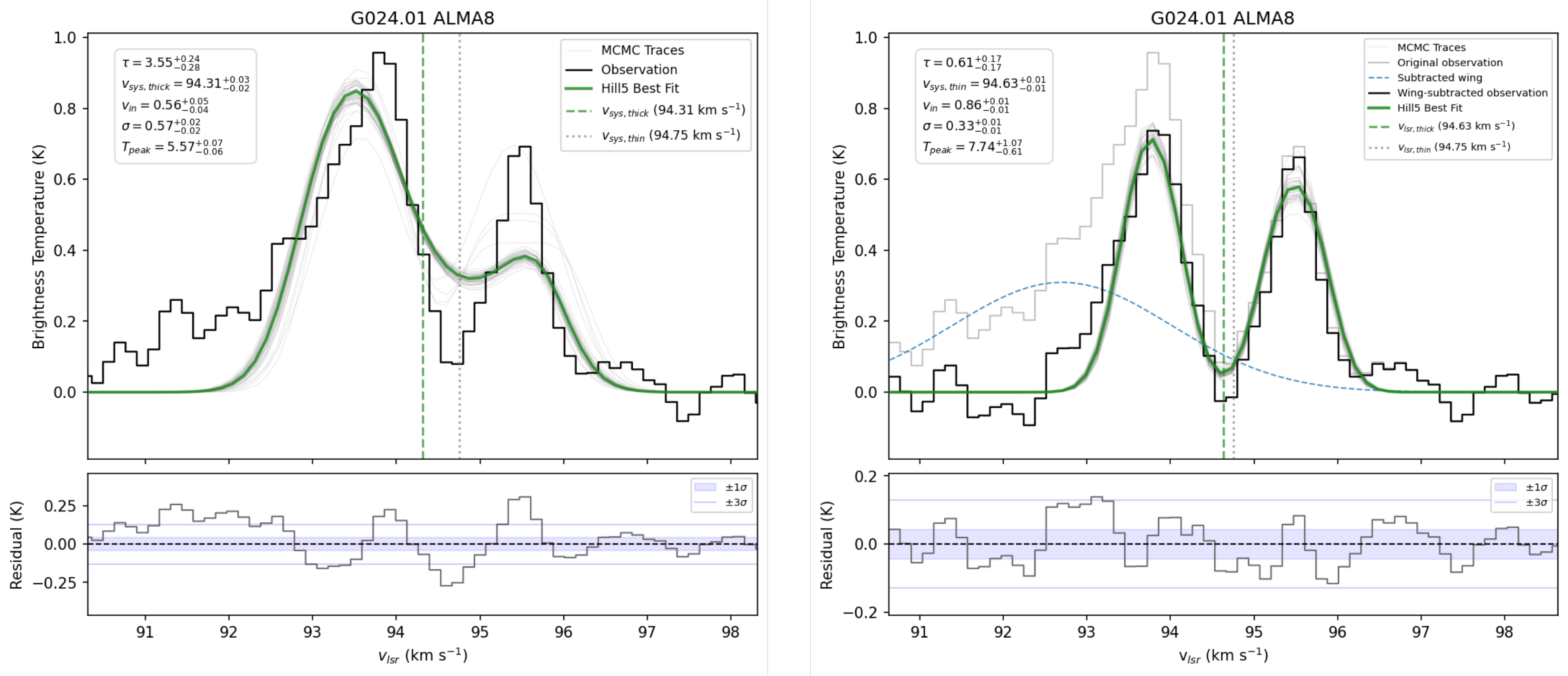}
     \caption{Example of the wing-subtraction procedure for the core G024.01 ALMA8. \textit{Left}: Hill5 model fit (green) to the original spectrum, as same style as Figure~\ref{fig:ex_profiles}. Thin gray lines show the posteriors of MCMC fitting. \textit{Right}: Hill5 model fit (green) to the wing-subtracted spectrum (black). The broad, outflow-associated wing component subtracted before refitting is shown as a dashed light-blue line. The profile before the subtraction is shown as a light gray line. }
    \label{fig:wingfit}
\end{figure}

For a subset of cores, the HNC line profile exhibits a broad wing or high-velocity tail extending toward the red- or blueshifted side, naturally attributed to high-velocity outflowing gas rather than to
the infall motion targeted by the Hill5 model. Such a broad component can bias the Hill5 fit, producing residuals that exceed the 3$\sigma$ threshold adopted as our success criterion (Section~\ref{sec:result}). Following the approach of \citet{Morii25}, we identify these cases as cores for which the first-pass Hill5 fit yields residuals larger than 3$\sigma$ and for which visual inspection reveals a clear broad wing or tail component at high velocity (e.g, the left panel of Figure~\ref{fig:wingfit}). 

For these cores, we decompose the observed spectrum into three Gaussian components via a least-squares fit: two narrow components associated with the blue- and redshifted infall peaks, and one broad, low-amplitude component that captures the high-velocity wing or tail attributed to outflow contamination. The middle panel of Figure~\ref{fig:wingfit} shows an example of this decomposition: the broad component (Gaussian~2) accounts for the excess emission extending beyond the two narrow peaks associated with infall.

We applied this three-component decomposition to 27 cores that satisfied the criteria above. After subtracting the broad component from each spectrum, we re-applied the Hill5 model fitting to the residual, wing-subtracted spectrum, following the same MCMC procedure described in Section~\ref{sec:result}. 
This second-pass fitting succeeded, by the same 3$\sigma$ residual criterion, for 14 cores.  The right panel of Figure~\ref{fig:wingfit} shows the resulting Hill5 fit after wing subtraction for the same example core. The remaining 15 cores did not yield a satisfactory fit, even after wing subtraction, and were excluded from the final sample. 

Thus, of the 70 successful Hill5 fits presented in this paper, 14 were obtained only after this wing/tail subtraction procedure, while the remaining 56 were fit directly from the original, unmodified spectra. 

\section{Additional Table} \label{sec:appendix-table}
Table~\ref{tab:parameters2} lists the parameters used in Figure~\ref{fig:Mdot_Vin_evol}, \ref{fig:Vin_phy}, \ref{fig:tin_tff}, and \ref{fig:Mdot_M_avir}. 
Full version is available online as a machine readable table. 

\begingroup
\renewcommand{\arraystretch}{1.15}
\begin{deluxetable*}{lllcccccccc}
\tablecaption{Summary of Physical Properties Used in the Figures\tablenotemark{*} \label{tab:parameters2}} 
\tablehead{
\colhead{Region} & \colhead{ID} &  \colhead{Evolutionary} & \colhead{Mass} & \colhead{$\dot{M}$} & \colhead{$v_{\rm sys, thin}$} & \colhead{$\sigma_{\rm nt}$} & \colhead{$\mathrm{M}$} & {$\alpha_{\rm vir}$} & \colhead{$t_{\rm ff}$} & \colhead{$t_{\rm in}$}\\
 \colhead{} & \colhead{}  & \colhead{Stage} & \colhead{($M_\odot$)} & \colhead{($M_\odot$\,yr)} & \colhead{(km\,s$^{-1}$)} & \colhead{(km\,s$^{-1}$)} & \colhead{} & \colhead{} & \colhead{10$^4$\,yr} & \colhead{10$^4$\,yr} 
}
\startdata
G016.97 &        4 & prestellar &   $0.32 \pm 0.17$ &    $1.30_{-0.74}^{+0.70}$ &  40.41 &      0.34 &  5.72 &  2.31 & $1.44_{-0.43}^{+0.43}$ &  $0.75_{-0.09}^{+0.16}$ \\
G016.97 &       11 & prestellar &   $0.60 \pm 0.31$ &    $0.55_{-0.29}^{+0.29}$ &  40.88 &      0.44 &  7.46 &  5.37 & $5.09_{-1.53}^{+1.53}$ &  $3.23_{-0.34}^{+0.34}$ \\
G018.80 &        7 & prestellar &     $2.7 \pm 1.3$ &    $2.93_{-1.45}^{+1.45}$ &  65.44 &      0.33 &  5.38 &  0.92 & $3.23_{-0.92}^{+0.92}$ &  $2.80_{-0.28}^{+0.28}$ \\
G018.93 &        7 & prestellar &   $1.12 \pm 0.56$ &    $0.72_{-0.36}^{+0.36}$ &  45.14 &      0.21 &  2.83 &  1.77 & $3.24_{-0.95}^{+0.95}$ &  $4.69_{-0.47}^{+0.47}$ \\
G022.69 &        4 & prestellar &     $2.4 \pm 1.1$ &    $0.80_{-0.53}^{+1.05}$ &  77.30 &      0.15 &  2.08 &  0.63 & $2.60_{-0.72}^{+0.72}$ & $8.83_{-10.84}^{+4.14}$ \\
G022.69 &       10 & prestellar &   $0.90 \pm 0.42$ &    $0.37_{-0.31}^{+0.18}$ &  76.88 &      0.40 &  5.66 &  4.01 & $4.74_{-1.32}^{+1.32}$ &  $7.19_{-0.99}^{+4.98}$ \\
G023.47 &        3 &    outflow &     $8.0 \pm 3.8$ & $23.46_{-11.50}^{+11.50}$ &  88.24 &      0.74 & 11.94 &  0.94 & $1.28_{-0.36}^{+0.36}$ &  $1.02_{-0.10}^{+0.10}$ \\
G023.47 &       13 & prestellar &   $0.79 \pm 0.38$ &    $0.85_{-0.42}^{+0.42}$ &  86.98 &      0.34 &  5.48 &  2.06 & $2.77_{-0.78}^{+0.78}$ &  $2.80_{-0.29}^{+0.29}$ \\
G023.47 &       17 &       warm &   $0.41 \pm 0.20$ &    $1.21_{-0.59}^{+0.59}$ &  88.15 &      0.92 & 14.92 & 17.96 & $2.94_{-0.83}^{+0.83}$ &  $1.01_{-0.10}^{+0.10}$ \\
G024.01 &        6 &       warm &   $1.57 \pm 0.76$ &    $2.73_{-1.35}^{+1.35}$ &  94.75 &      0.73 & 12.37 &  3.52 & $1.91_{-0.54}^{+0.54}$ &  $1.72_{-0.17}^{+0.17}$ \\
G024.01 &        8 &       warm &   $1.68 \pm 0.81$ &    $3.12_{-1.54}^{+1.54}$ &  94.75 &      0.18 &  3.01 &  0.48 & $2.17_{-0.62}^{+0.62}$ &  $1.62_{-0.16}^{+0.16}$ \\
G024.52 &        2 &    outflow &     $5.4 \pm 2.6$ &   $13.76_{-6.78}^{+6.78}$ &  91.16 &      0.86 & 14.28 &  0.93 & $0.57_{-0.16}^{+0.16}$ &  $1.19_{-0.12}^{+0.12}$ \\
G024.52 &        6 &    outflow &     $4.5 \pm 2.2$ &    $2.73_{-1.34}^{+1.34}$ &  92.52 &      0.39 &  6.53 &  0.65 & $2.12_{-0.60}^{+0.60}$ &  $4.91_{-0.50}^{+0.50}$ \\
G024.52 &        7 &       warm &     $3.8 \pm 1.8$ &    $3.79_{-1.95}^{+1.93}$ &  92.05 &      0.50 &  8.36 &  1.14 & $2.30_{-0.65}^{+0.65}$ &  $3.01_{-0.49}^{+0.53}$ \\
G024.52 &        9 &       warm &     $2.9 \pm 1.4$ &    $0.46_{-0.29}^{+0.28}$ &  91.41 &      0.38 &  6.29 &  0.97 & $2.72_{-0.77}^{+0.77}$ & $18.75_{-6.84}^{+7.23}$ \\
G024.52 &       11 &       warm &   $1.00 \pm 0.48$ &    $2.13_{-1.05}^{+1.05}$ &  93.25 &      0.15 &  2.47 &  0.64 & $2.26_{-0.64}^{+0.64}$ &  $1.41_{-0.15}^{+0.15}$ \\
G024.52 &       13 & prestellar &   $0.59 \pm 0.29$ &    $0.89_{-0.45}^{+0.60}$ &  93.38 &      0.38 &  6.30 &  2.41 & $2.16_{-0.61}^{+0.61}$ &  $1.99_{-0.91}^{+0.27}$ \\
G024.52 &       19 & prestellar &   $0.73 \pm 0.35$ &    $1.16_{-0.68}^{+0.57}$ &  93.14 &      0.47 &  7.81 &  3.87 & $3.25_{-0.92}^{+0.92}$ &  $1.88_{-0.22}^{+0.64}$ \\
G025.16 &        8 &    outflow &     $2.8 \pm 1.4$ &    $4.58_{-2.30}^{+2.30}$ &  64.68 &      0.47 &  7.97 &  1.34 & $2.48_{-0.72}^{+0.72}$ &  $1.83_{-0.18}^{+0.18}$ \\
G025.16 &        9 &       warm &   $0.96 \pm 0.47$ &    $0.45_{-0.25}^{+0.27}$ &  64.00 &      0.50 &  8.36 &  2.39 & $1.80_{-0.52}^{+0.52}$ &  $6.44_{-2.33}^{+1.85}$ \\
G025.16 &       14 &       warm &   $0.96 \pm 0.47$ &    $1.39_{-0.70}^{+0.70}$ &  64.16 &      0.21 &  3.54 &  0.85 & $2.23_{-0.64}^{+0.64}$ &  $2.07_{-0.22}^{+0.22}$ \\
G028.54 &        2 &    outflow &     $8.3 \pm 4.1$ &    $5.53_{-2.77}^{+2.77}$ &  86.03 &      0.22 &  3.64 &  0.28 & $3.00_{-0.86}^{+0.86}$ &  $4.52_{-0.46}^{+0.45}$ \\
G028.54 &        4 &       warm &     $3.8 \pm 1.9$ &    $3.41_{-1.70}^{+1.70}$ &  85.80 &      0.46 &  7.41 &  1.19 & $2.99_{-0.86}^{+0.86}$ &  $3.39_{-0.34}^{+0.34}$ \\
G028.54 &        5 &       warm &     $6.5 \pm 3.2$ &    $4.78_{-2.39}^{+2.39}$ &  86.58 &      0.34 &  5.55 &  0.68 & $4.14_{-1.19}^{+1.19}$ &  $4.06_{-0.41}^{+0.41}$ \\
G028.54 &        9 & prestellar &   $0.68 \pm 0.33$ &    $1.34_{-0.67}^{+0.67}$ &  87.02 &      0.43 &  6.92 &  3.02 & $2.51_{-0.72}^{+0.72}$ &  $1.51_{-0.16}^{+0.16}$ \\
G028.54 &       15 & prestellar &   $0.65 \pm 0.32$ &    $0.51_{-0.37}^{+0.57}$ &  86.12 &      0.66 & 10.74 &  8.13 & $3.54_{-1.02}^{+1.02}$ &  $3.87_{-3.90}^{+2.09}$ \\
G028.56 &        2 &    outflow &       $26 \pm 13$ & $51.34_{-25.96}^{+25.96}$ &  87.73 &      0.53 &  9.04 &  0.20 & $0.99_{-0.29}^{+0.29}$ &  $1.50_{-0.15}^{+0.15}$ \\
G028.92 &        5 & prestellar &   $1.62 \pm 0.78$ &    $0.92_{-0.52}^{+0.51}$ &  96.14 &      0.29 &  4.51 &  1.17 & $2.83_{-0.80}^{+0.80}$ &  $5.28_{-1.46}^{+1.56}$ \\
G030.70 &        3 &       warm &     $3.4 \pm 1.7$ &    $3.30_{-1.66}^{+1.66}$ &  93.12 &      0.24 &  3.74 &  0.54 & $2.65_{-0.76}^{+0.76}$ &  $3.08_{-0.31}^{+0.31}$ \\
G030.70 &        4 &       warm &     $3.6 \pm 1.8$ &    $1.99_{-1.22}^{+1.10}$ &  92.87 &      0.31 &  4.76 &  0.85 & $3.79_{-1.09}^{+1.09}$ &  $5.47_{-1.38}^{+2.01}$ \\
G030.70 &       13 &       warm &   $0.44 \pm 0.22$ &    $0.52_{-0.28}^{+0.30}$ &  93.04 &      0.29 &  4.56 &  2.67 & $2.77_{-0.80}^{+0.80}$ &  $2.58_{-0.78}^{+0.60}$ \\
G030.70 &       14 & prestellar &   $0.57 \pm 0.28$ &    $0.68_{-0.36}^{+0.38}$ &  93.18 &      0.11 &  1.77 &  1.22 & $3.08_{-0.89}^{+0.89}$ &  $2.52_{-0.69}^{+0.44}$ \\
G030.70 &       15 & prestellar &   $0.53 \pm 0.26$ &    $0.62_{-0.44}^{+0.38}$ &  93.07 &      0.36 &  5.56 &  3.54 & $3.46_{-1.00}^{+1.00}$ &  $2.55_{-0.90}^{+1.29}$ \\
\enddata
\tablenotetext{*}{Full version is available online as machine-readable table. }
\end{deluxetable*}
\endgroup

\bibliography{references}
\bibliographystyle{aasjournalv7}

\end{document}